\documentclass[twocolumn,preprintnumbers,amsmath,amssymb,floatfix,superscriptaddress,nofootinbib]{revtex4}

\usepackage{amsmath,hyperref,amssymb}
\usepackage{graphicx}
\usepackage{dcolumn}
\usepackage{bm}
\usepackage{verbatim}
\usepackage{tabularx}
\usepackage{slashed}
\usepackage{cancel}
\usepackage[boxsize=2em,aligntableaux=center]{ytableau}
\usepackage{float}
\usepackage{color}
\usepackage{xcolor}
\usepackage{enumerate}

\def\be{\begin{equation}}
\def\ee{\end{equation}}
\def\bea{\begin{eqnarray}}
\def\eea{\end{eqnarray}}

\begin{document}

\vspace*{-30mm}

\title{Discovering supernova-produced dark matter with directional detectors}

\author{Elisabetta Baracchini}
\affiliation{Gran Sasso Science Institute, I-67100, L'Aquila, Italy}
\affiliation{Istituto Nazionale di Fisica Nucleare Laboratori Nazionali del Gran Sasso, I-67100 Assergi, Italy}
\author{William DeRocco}
\affiliation{Stanford Institute for Theoretical Physics, \\
Stanford University, Stanford, CA 94305, USA}
\author{Giorgio Dho}
\affiliation{Gran Sasso Science Institute, I-67100, L'Aquila, Italy}
\affiliation{Istituto Nazionale di Fisica Nucleare Laboratori Nazionali del Gran Sasso, I-67100 Assergi, Italy}

\vspace*{1cm}

\begin{abstract} 

Supernovae can produce vast fluxes of new particles with masses on the MeV scale, a mass scale of interest for  models of light dark matter. When these new particles become diffusively trapped within the supernova, the escaping flux will emerge semirelativistic with an order-one spread in velocities. As a result, overlapping emissions from Galactic supernovae will produce an overall flux of these particles at Earth that is approximately constant in time. However, this flux is highly anisotropic and is steeply peaked towards the Galactic center. This is in contrast with the cosmological abundance of a WIMP-like dark matter which, due to the rotation of the Galaxy, appears to come from the direction of the Cygnus constellation. In this paper, we demonstrate the need for a directional detector to efficiently discriminate between a signal from a cold cosmological abundance of GeV-scale WIMPs and a signal from a hot population of supernova-produced MeV-scale dark matter.

\end{abstract}

\maketitle

\tableofcontents

\section{Introduction}
Several astrophysical measurements (including those of the Cosmic Microwave Background, cluster and galaxy rotations, gravitational lensing, and Big Bang nucleosynthesis) indicate that the majority of the matter in the Universe is cold and dark (i.e. non-luminous and non-absorbing)~\cite{Bertone:2004pz}. Taken together, these observations suggest the existence of at least one quasi-stable dark matter (DM) particle that is not predicted by the Standard Model of particle physics. 
The nature of dark matter is however still a mystery: deciphering its essence is one of the most compelling tasks for fundamental physics. 

Weakly Interacting Massive Particles (WIMPs) are well-motivated dark matter candidates, independently predicted by Standard Model extensions and Big Bang cosmology. A stable, weakly-interacting particle at the GeV scale in thermal equilibrium with the early Universe would reproduce the observed relic dark matter density. Particles of this mass can appear in many theories, for example, as the lightest superpartner in supersymmetric models that conserve  R-parity~\cite{Jungman:1995df} or as the lightest Kaluza-Klein particle in Universal Extra Dimensions models~\cite{Hooper:2007qk}. 

For these reasons, most direct detection approaches have concentrated up to now on experimentally detecting WIMPs by looking for a very low energy (1-100 keV) nuclear (or electron) recoil due to a DM particle undergoing an elastic scatter in the active volume of a detector. The expected WIMP scattering direction is the result of the Earth's relative motion with respect to the Galactic halo, which is thought to contain a high concentration of DM due to measurements of Galactic rotation curves. This motion produces an apparent WIMP wind coming from the Cygnus constellation on Earth, with a change in direction of $\sim$ 90$^{\circ}$ for every 12 sidereal hours due to Earth's axis orientation with respect to the DM wind \cite{Spergel:1987kx}. Typical DM detectors aim at measuring only the energy deposited by the scattered nuclei in the target medium, while only a few, mostly in the research and development stage, are trying to additionally observe the recoil track orientation that encodes the WIMP arrival direction. It has been already shown that the determination of the incoming direction of the WIMP particle can provide a correlation with an astrophysical source that backgrounds cannot mimic and therefore offers a unique key for a positive, unambiguous identification of a DM signal \cite{Mayet:2016zxu}.

WIMPs are, however, not the only paradigm that can explain dark matter. As the parameter space for WIMPs has become increasingly constrained by direct detection experiments, theoretical and experimental focus has broadened to alternative candidates. One such alternative paradigm is known as \textit{light dark matter}, which is taken to mean dark matter candidates with masses below a GeV~\cite{Izaguirre:2015yja}. By virtue of their low mass, the energy deposition by these particles would be too low to have been observed in typical direct detection experiments. An interesting effect of these low masses is that these particles can be produced abundantly in astrophysical sources such as supernovae. As was shown recently, there are regions of parameter space in which a hot Galactic population of dark matter can be produced by supernovae with momenta large enough to be detected in existing and proposed WIMP detectors~\cite{DeRocco:2019jti}.

By virtue of the comparable momenta between a non-relativistic WIMP and semi-relativistic sub-GeV particle, light dark matter produced in supernovae (henceforth SNDM) will induce an energy response in DM detectors that is very difficult to discriminate from that of a WIMP. In this paper we demonstrate how directional detection provides a unique handle in discriminating not only different WIMP models, but also between WIMP-like DM and an SNDM signal. Due to the high degree of anisotropy in the angular distribution of recoils induced by SNDM (whose production is strongly peaked in the Galactic center), directional DM detectors can serve as a critical tool for discriminating the two scenarios.

The layout of this paper is as follows. In Section~\ref{sec:status}, we discuss the current state of directional and non-directional DM detectors. In Section~\ref{sec:theory}, we then review the physics behind the production of an SNDM population in the Galaxy. Section~\ref{sec:kinematic} describes the kinematics for both WIMP and SNDM scattering in a generic detector and provides analytic formulae for recoil spectra. In Section~\ref{sec:toy}, we introduce a fiducial toy experimental setup which we use to showcase the potential sensitivity of directional detectors to SNDM. Finally, we display the results of our analysis of the discriminatory capabilities of these toy experiments in Section~\ref{sec:results} and summarize our conclusions in Section~\ref{sec:conc}.

\section{Current status of direct DM searches}\label{sec:status}

In this section, we review the current status of directional dark matter detection.

Historically, WIMP detectors have focused mainly on using noble liquids (e.g. argon and xenon), in order to exploit the high density and impressive scintillating properties of these media. Liquid Xenon-based Time Projection Chamber (TPC) experiments have produced the most stringent upper limits to date on DM scattering in the 10-1000 GeV WIMP mass range~\cite{Aprile:2019dbj}.

However, these experiments are only sensitive to the energy of the recoiling nuclei and do not possess directional sensitivity. As a result of this, the irreducible background  induced by $^{8}$B solar neutrinos at $10^{-44}-10^{-45}$ cm$^2$~\cite{Billard:2013qya} contributing to the so-called ``neutrino floor'' will strongly limit the sensitivity of present and future non-directional DM experiments of this kind below a $\sim10$ GeV WIMP mass for both spin-independent and spin-dependent couplings. 

Alternative approaches have been proposed to explore the low-mass region above the neutrino floor between 1 and 10 GeV WIMP masses, mainly based on cryogenic bolometers (e.g. SuperCDMS~\cite{Agnese:2013rvf} and CRESST~\cite{Abdelhameed:2019hmk}), or high-pressure neon-based TPCs with one channel readouts (e.g. NEWS-G~\cite{Arnaud:2017bjh}), however these too will be limited by the solar neutrino background.

Directional detection, in contrast, not only provides an unique key for a positive identification of a DM signal, but would also allow experiments to probe below the neutrino floor~\cite{Billard:2013qya}. The main experimental challenge of directional DM experiments in the case of gaseous media is to instrument a very large volume with high spatial granularity to image $\mathcal{O}$(mm) tracks, while still being able to control its backgrounds. The required spatial resolution is ultimately set by the density of the target material, since it determines the characteristic length of a WIMP-induced recoil, which in turn fixes the minimum energy threshold for which a direction can be distinguished. 

Critically, both the track length and the ability to infer the real initial WIMP direction are highly affected by the density of the target material, as is discussed in Ref.~\cite{Couturier:2016isu}. For example, in solid targets, a nucleus recoiling with $\leq$ 100 keV will travel less than 100 nm, while in TPCs with low density gaseous targets it will travel up to 1 mm at 50 keV recoil. Additionally, once the recoiling nucleus encounters its first elastic scattering off another nucleus in the sensitive medium, the knowledge of the initial recoil direction starts to get diluted and lost, a process that happens with a far greater rate in solid rather than gaseous targets. The subsequent scatters generate a non-negligible, irreducible angular dispersion, which is usually referred to as the ``straggling'' effect, and that can measure up to 25$^{\circ}$ (rms) at 10 keV in pure CF$_{4}$ gas~\cite{Billard_thesis}. 

In recent years there has been a resurgence of interest in community towards in the possibility of introducing recoil direction sensitivity into the field~\cite{Battaglieri:2017aum}. In parallel, significant progresses have been made in readout technologies enabling high precision tracking in gas TPCs down to low energy~\cite{Battat:2016pap}. As a consequence of these developments, a new international proto-collaboration has been formed, called CYGNUS~\cite{Miuchi:2020igv, Vahsen:2020pzb}, with the aim of developing a global network of recoil-sensitive TPCs to be used for a directional dark matter search. The key features of the proposed experiment are a modular design of recoil-sensitive TPCs filled with a Helium-Fluorine based gas mixture with installation in multiple underground sites to minimize location systematics and improve sensitivity (Boulby Underground Laboratory, Laboratori Nazionali del Gran Sasso, Kamioka Underground Mine, Stawell Mine).

The choice of the optimal detection technique and exact gas mixture is still under evaluation and study, and for this reason six different groups inside the collaboration are working on $\mathcal{O}(1)$ m$^3$ demonstrator projects, with a large variety of readouts and amplification techniques~\cite{Ezeribe:2019tln, Hashimoto:2020xly, Baracchini:2020phn, Vahsen:2014fba, Phan:2017sep, Santos:2013hpa} and gases with different properties~\cite{Martoff:2000wi,Snowden-Ifft:2014taa, Phan:2016veo, Pinci:2019hhw} being tested at the moment. These ongoing experiments inform our choice of fiducial parameters in the toy experimental setup we introduce in Section~\ref{sec:toy}.

As will be shown in the remainder of this paper, directional dark matter detectors have the ability to discriminate between various models of dark matter with far greater efficiency than their non-directional counterparts.

\section{The diffuse Galactic DM flux from supernovae}\label{sec:theory}

\begin{figure*}
  \centering
  \includegraphics[width=0.9\textwidth]{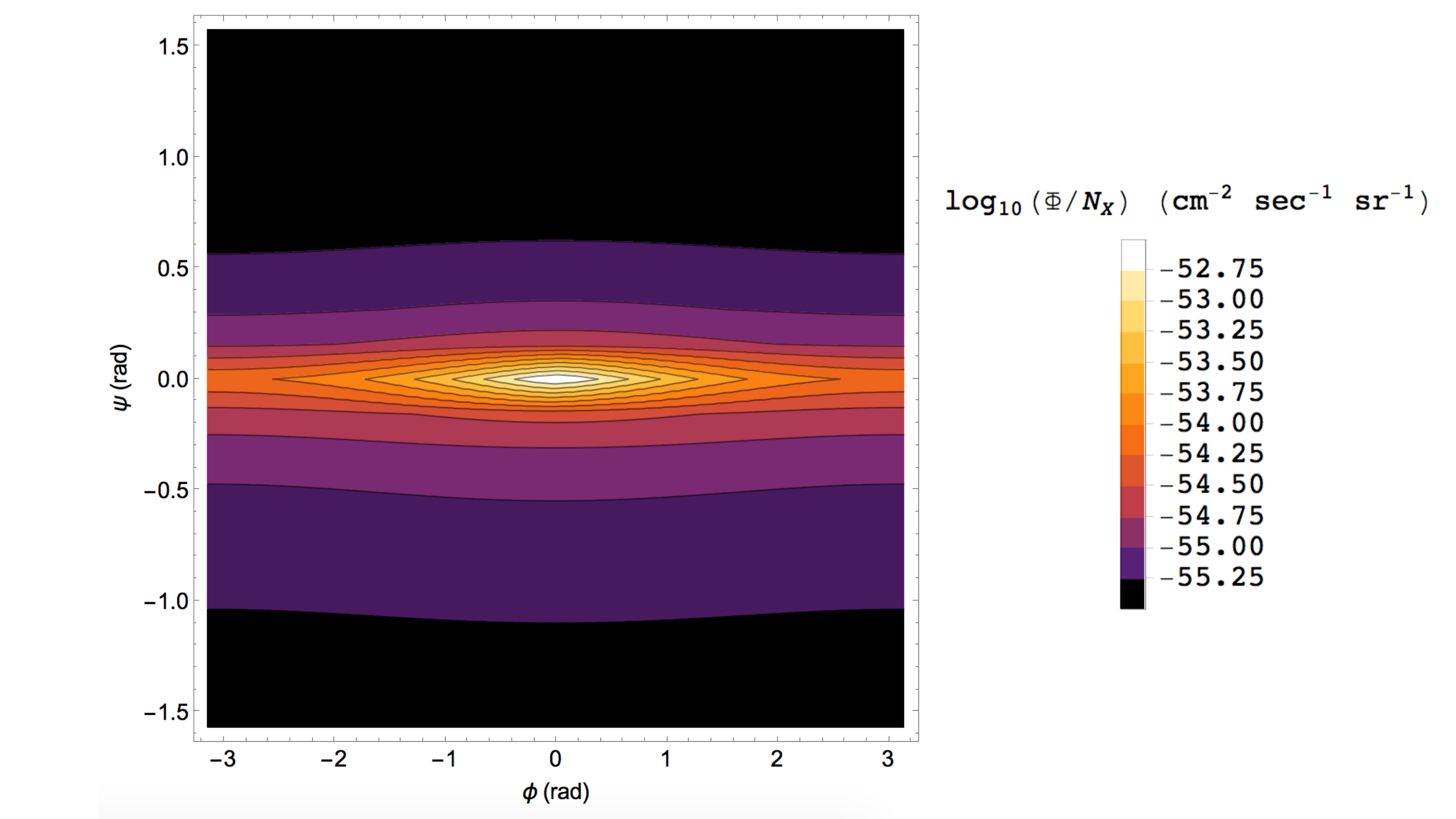}
  \caption{Sky map for the flux of light DM produced by Galactic SNe. The scale has been normalized by $N_X$, the total number of DM particles produced in a single supernova. It is evident that the increased rate of SNe in the Galactic center results in a large flux from that region. Note that the expected flux looking directly out of the plane of the Galaxy is four orders of magnitude smaller than the flux coming from the Galactic center.
     \label{Fig: skymap}}
\end{figure*}

In this section, we show that the production by supernovae of an approximately constant-in-time but highly anisotropic flux of semirelativistic DM is a generic feature of DM models on the MeV scale with suitably strong coupling to the Standard Model.

As was first pointed out in Ref.~\cite{DeRocco:2019jti}, Galactic supernovae can produce a flux of hot MeV-scale DM at Earth that is roughly constant in time. In the following paragraphs, we will summarize the results of that paper. Note that while Ref.~\cite{DeRocco:2019jti} treats a specific example model, this is in fact a general feature of models with new degrees of freedom on the MeV scale. Since the temperature of a core-collapse supernova (SN) can reach upwards of 30 MeV, supernovae will produce a large thermal flux of particles with masses up to hundreds of MeV (at which point the flux becomes heavily Boltzmann-suppressed). If the new degree of freedom is coupled sufficiently strongly to the Standard Model (SM) that it becomes diffusively trapped within the protoneutron star (PNS), it will stay in thermal contact with the SM bath out to some radius (the ``energy sphere'' $r_E$). Akin to the SN neutrinos, the DM flux will then approximate blackbody emission from this sphere with a temperature set by the local temperature at the energy sphere. (See Ref.~\cite{DeRocco:2019jti} for further details.)

For sufficiently massive particles ($m_X > T(r_E)$, with $m_X$ the mass of the new particle), the escaping flux will be semirelativistic with a velocity distribution approximately Maxwell-Boltzmann at the temperature of the energy sphere. The Maxwell-Boltzmann distribution exhibits a roughly order-one spread in velocities (i.e. $\sim75\%$ of the distribution lies between $v = 0.5\bar{v}$ and $v = 1.5\bar{v}$ where $\bar{v}\approx \sqrt{T/m_X}$ is the average speed). This spread in velocities leads to a spread in arrival times of these particles at Earth. For a supernova a distance $d$ away, the spread in arrival time is $\Delta t \approx (d/\bar{v})\delta v$ with $\delta v \equiv \frac{\Delta v}{v} \approx 1$.

Consider a supernova that occurs in the center of our Galaxy ($\sim 3000$ light years away). The DM flux will be produced over $\sim 10$ seconds (the cooling timescale of the PNS), but the arrival on Earth of the bulk of the flux will take place over $\sim 3000(c/\bar{v})$ years. Since Type II supernovae are predicted to occur in the Galaxy at a rate of roughly 2 per century~\cite{Beacom:2010kk}, the emissions of $\gtrsim 100$ SN will all be overlapping at Earth at any  given time. These overlapping emissions produce what we call the ``diffuse galactic flux'' of hot dark matter produced in supernovae (SNDM).

While we use the term ``diffuse'' to indicate that this flux is approximately constant in time, it is \textit{not} isotropic. It is instead very strongly peaked towards the center of the Galaxy, which is where the majority of supernovae take place. To quantify this, we use the double-exponential profile of Adams et al.~\cite{Adams:2013ana} for the core-collapse SN density rate in our galaxy:
\be
\frac{dn_{SN}}{dt} = A e^{-r/R_d} e^{-|z|/H}
\ee
with $R$ the galactocentric radius and $z$ the height above the Galactic mid-plane. Earth sits at $R_E = 8.7$ kpc and $z_E = 24$ pc. We choose to adopt the parameter values Adams et al. provide for Type II SN: $R_d = 2.9$ kpc, $H =$ 95 pc. We normalize the SN rate to two SN per century.

We compute the flux along a given line of sight by performing the following integration:
\begin{multline}
\Phi(\psi, \phi) = \\
N_{X}\int_0^{\infty} \frac{dn_{SN}}{dt}\biggm\lvert_{\left\{\begin{subarray}\\r = \sqrt{r_E^2+(l\cos\psi)^2-2r_E(l\cos\psi\cos\phi)},\\
z = z_E + l\sin\psi\end{subarray}\right\}}~dl
\end{multline}
with $\psi$ and $\phi$ the polar and azimuthal angles respectively and $N_X$ the total number of SNDM particles emitted by a single SN. In Fig.~\ref{Fig: skymap}, we display the result of this computation with $N_X$ divided out, as it simply adjusts the normalization.

We see that the flux of DM due to supernovae is strongly peaked towards the Galactic center. The flux from this region exceeds that of the flux coming from directly out of the plane by four orders of magnitude.

Note that for this entire discussion, we have focused on supernovae occurring within our own galaxy. It is natural to ask about whether there is an isotropic contribution to this flux due to extragalactic supernovae, which is, for example, how the diffuse supernova neutrino background is formed~\cite{Beacom:2010kk}. However, the neutrinos are all traveling at $c$, which means that they experience no time-spreading effect. For massive particles, the time-spreading effect becomes so large on extragalactic scales that the corresponding flux, even integrated out to high redshift, is subdominant to the Galactic flux. We can perform another line of sight integral as specified by Ref.~\cite{Beacom:2010kk}:
\be
\Phi = N_X\int_0^{\infty} R_{SN}(z)\left|\frac{dt}{dz}\right|~dz
\ee
where $R_{SN}(z)$ is the redshift-dependent Type II supernova rate (taken from~\cite{Beacom:2010kk}) and $(\frac{dt}{dz})^{-1} = H_0 (1+z) \sqrt{ \Omega_{\Lambda} + \Omega_{m} (1+z)^3}$. Performing this integral and dividing by $N_X$ as above, we compute $\log_{10}(\Phi/N_X) \approx -56.3$. Note that this is an order of magnitude below even the weakest line of sight for the Galactic contribution (directly out of the plane). As a result, we choose to ignore the isotropic extragalactic contribution for the remainder of this paper.

To summarize the preceding discussion, the production of an anisotropic constant flux of high-momentum particles by Galactic SNe is a generic feature of any DM model with mass $\mathcal{O}(10)-\mathcal{O}(100)$ MeV and coupling to the SM sufficient to diffusively trap the particle within the PNS out to $\sim 10-20$ km.

\section{Scattering kinematics and recoil spectra}\label{sec:kinematic}

In this section, we will discuss the scattering kinematics of both SNDM and WIMPs in nuclear recoil experiments. We will begin with a simple explanation of the difficulties in discrimination using energy spectra (Sec.~\ref{sec:qualkin}), then focus on a specific model of SNDM (Sec.~\ref{sec:toymodel}) for which we can explicitly produce an analytic formula for the scattering kinematics (Sec.~\ref{sec:sndmrecoil}). This will then be compared to the same formula for WIMPs (Sec.~\ref{sec:WIMPrecoil}).

\subsection{Comparing SNDM and WIMP scattering}
\label{sec:qualkin}

In the standard WIMP picture, WIMPs have masses of $\mathcal{O}(10) - \mathcal{O}(100)$ GeV~\cite{Schumann:2019eaa}, three orders of magnitude greater than the SNDM discussed in the previous section. However, direct detection experiments built to target WIMPs are actually sensitive to SNDM as well. This is due to the simple fact that the cosmological abundance of WIMPs travel at the Galactic virial velocity ($v\approx 10^{-3} c$) while the SNDM travels at some order-one fraction of $c$. The semi-relativistic velocity of the SNDM compensates for its lighter mass and results in a similar momentum to a cold GeV-scale WIMP.

We will compute the momentum of a recoiling target nucleus in a DM detector with no assumptions that the momentum transfer is non-relativistic. The recoil momentum of the target is
\be
\label{eq:momentum}
|\vec{k}| = \frac{2 m_A (\sqrt{p_0^2 + m_X^2} + m_A) p_0 \cos\theta_r}{(\sqrt{p_0^2 + m_X^2} + m_A)^2 - p_0^2 \cos^2 \theta_r}
\ee
with $p_0$ the 3-momentum of the incident DM, $\theta_r$ the angle between the incident DM velocity and recoiling nucleus in the lab frame, and $m_X$ and $m_A$ the mass of the DM and target respectively. 

Let us consider the case of a GeV-scale WIMP striking a nucleus. In this case, we have a hierarchy of scales $m_X, m_A \gg p_0$. Treating $p_0$ as small gives the expression 
\be
|\vec{k}^{\text{WIMP}}_{\text{nuc}}| \approx 2 p_0 \cos\theta_r \left(\frac{m_A}{m_A + m_X}\right).
\ee
It is clear that this reduces to the usual expression for the recoil momentum imparted by a WIMP~\cite{Gondolo:2002np}: $|\vec{k}| = 2 v_0 \mu \cos\theta_r$ with $\mu = m_X m_A / (m_X + m_A)$ the reduced mass of the system.

In the case of MeV-scale DM traveling semirelativistically, we have $m_A \gg m_X, p_0$. Hence, we find that
\be
|\vec{k}^{\text{SNDM}}_{\text{nuc}}| \approx 2 p_0 \cos\theta_r.
\ee
We see immediately that we can write $p_0 = \gamma_0 m_X v_0$ with $\gamma_0$ the Lorentz factor and get $|\vec{k}| = 2 E_0 v_0 \cos\theta_r$, which is identical to the case of the WIMP up to $\mu \rightarrow E_0$.

This poses a significant challenge. One would wish to be able to discriminate between a cold GeV-scale WIMP and hot SNDM if some signal were to be detected. Unfortunately, since $\mu v_0$ in the WIMP scenario and $E_0 v_0$ in the SNDM scenario can be of comparable order, even if one were able to generate large statistics on the recoil energy distribution in a detector, it would be very difficult to discriminate between the two models. Differences will show up in the recoil spectrum due to differences in the shape of the incoming momentum spectrum, but both are approximately Boltzmann and there is considerable uncertainty on the WIMP velocity distribution~\cite{RevModPhys.85.1561} and momentum spectrum of the SNDM that could limit discrimination. We will quantify just how difficult it is to discriminate these models with energetic information alone in Section~\ref{sec:results}.\footnote{One could also attempt to discriminate using electron recoils or annual modulation. See Appendix~\ref{app:elec} for a discussion of these cases.}

The best discriminator is the fact that, as will be explicitly shown in the following sections, the SNDM is highly anisotropic, with a steep peak towards the Galactic center. WIMPs, in contrast, appear to originate from the Cygnus constellation due to the rotation of our Galaxy. This means that the signals from SNDM and from a cosmological abundance of WIMPs would be \textit{perpendicular}. In this way, directional detection can allow us to discriminate these two populations with a very small number of events.

\subsection{SNDM example model}
\label{sec:toymodel}
In order to evaluate the expected signal from the SNDM in our fiducial experiment (introduced in Section~\ref{sec:toy}), we will focus on a specific example model of SNDM that produces the features discussed in Sec.~\ref{sec:theory}. (This is the same model as used in in Ref.~\cite{DeRocco:2019jti}.) 

Namely, it is a dark sector with a Dirac fermion coupled to the SM via the four-fermion operator
\be
\label{eq:interaction}
	\mathcal{L} \supset \frac{e \epsilon g_d}{\Lambda^2} \bar \chi \gamma_\mu \chi J^\mu_\text{em}
\ee
with $\chi$ the dark matter and $J^\mu_\text{em}$ the electromagnetic current of the SM. This can be viewed as the case in which the dark sector is coupled to the SM by a dark photon with mass $\Lambda$ and kinetic mixing $\epsilon$ and a dark charge $g_d$. We take $\Lambda \gg \mathcal{O}(10)$ MeV large such that it is not produced on-shell in the SN. We further parametrize the coupling to the SM by the variable $y\equiv \frac{\epsilon^2 g_d^2}{4\pi} \left(\frac{m_{\chi}}{\Lambda}\right)^4$~\cite{Izaguirre:2015yja}.

As a result of this choice of coupling term, the predominant interaction that keeps the dark sector thermally coupled to the SM is scattering with electrons. It is the radius at which this interaction ceases to become efficient that the temperature of the escaping DM is set (the ``energy sphere'' described in Sec.~\ref{sec:theory}). This is defined formally by finding the radius $r_E$ at which the optical depth for this interaction ($\tau_E$) is 2/3:
\be
\label{eq:opticaldepth}
\tau_E|_{r=r_E} \equiv \int_{r_E}^{\infty} \sqrt{\lambda_{\chi e}^{-1}(r)[\lambda_{\chi p}^{-1}(r) + \lambda_{\chi e}^{-1}(r)]}~dr = \frac{2}{3}
\ee
where
$\lambda_{\chi p}$
and 
$\lambda_{\chi e}$
are the mean free path for DM scattering with protons and electrons respectively.

This condition simply provides a mathematical prescription to determine at what radius the SNDM decouples from the Standard Model thermal bath. At radii $r < r_E$, the SNDM is undergoing rapid scatters off of electrons, which allows the SNDM population to stay in thermal equilibrium with the Standard Model. However, at radii $r > r_E$, the electron density has dropped to the point at which the average SNDM particle will escape to infinity without undergoing any more electron scatters. The optical depth computed in Eq.~\ref{eq:opticaldepth} is roughly the number of scatters expected for an SNDM particle to undergo as it escapes. For this reason, when it drops below unity (or, more formally, 2/3~\cite{Raffelt:2001kv}), this simply means that SNDM particles emitted at that radius will not exchange energy with the SM bath and will not be thermally coupled to the Standard Model. Instead, their temperature will be set at the last radius at which they \text{were} thermally-coupled to the SM, which is just the definition of the ``energy sphere.'' So, as stated previously, it is the local temperature of the SM bath at this energy sphere that sets the temperature of the escaping SNDM flux, which in turn sets the momentum spectrum of incident SNDM at an Earth-based detector.\footnote{Note that the proton scattering appears in the formula as the SNDM is still undergoing diffusive scatters off of the protons that do \textit{not} exchange energy even once the electron scatters have become inefficient.} For further details, see Ref.~\cite{DeRocco:2019jti}.

\subsection{SNDM recoil spectrum}
\label{sec:sndmrecoil}

Due to the high degree of anisotropy in the angular distribution of SNDM evident from Fig.~\ref{Fig: skymap}, we approximate the dark fermion flux as being produced from a point source at the Galactic center. Since the distribution of SNDM particles follows a Fermi-Dirac distribution
\be
\label{eq:sn_dirac}
    f(p)=\frac{A}{e^{\frac{\sqrt{p^2+m_X^2}}{T}}+1}
\ee 
where $p$ and $m_X$ are the SNDM momentum and mass respectively, and $T$ the temperature at the energy sphere, the differential nuclear recoil spectrum can be computed in the standard way (see, e.g., Eq. 16 of Ref.~\cite{Gondolo:2002np}):
\be
\label{eq:diff_cs}
\frac{dR}{dq^2~d\Omega}=\frac{N}{Nm_A}\int\frac{d\sigma}{dq^2d\Omega}nvf(\vec{v})d^3v
\ee
where $N$ is the number of nuclei in the detector, $m_A$ is the target mass, $d\Omega$ denotes an infinitesimal solid angle around the nuclear recoil direction $\hat{q}$, $n$ is the DM number density, and $f$($\vec{v}$) is the DM velocity distribution in the frame of the detector.

By using the relations $q^2=2m_A E$ (with $E$ the nuclear recoil energy) and $f(\vec{v})~d^3v=f(\vec{p})~d^3p$ and reabsorbing all the constant coefficients in the term $A'$, we can rewrite Eq.~\ref{eq:diff_cs} as:
\be
\label{eq:sn_drdedcos}
\frac{dR}{dE~d\cos\theta}=A' \int \delta(\cos\theta-\frac{q}{2p})S(q)\frac{p}{m_X}\frac{f(p)}{\sqrt{1+\frac{p^2}{m_X^2}}}~dp
\ee
where $S(q)$ is the square modulus of the nuclear form factor.

As it turns out, the effect of gravitational redshift on the SNDM particle energies during escape from the SN is not irrelevant~\cite{DeRocco:2019jti}. As a consequence, the SNDM momentum (denoted $p$) measured at the lab is related to the one at the protoneutron star's energy sphere (denoted $p_*$)  by $p_*^2=\sqrt{\frac{p^2+2\Phi m_X^2}{1-2\Phi}}$ with $\Phi\equiv \int_{r_E}^{\infty} \frac{m_{\text{enc}}(r)}{r}~dr$ where $m_{\text{enc}}(r)$ is the protoneutron star mass enclosed in a radius $r$. Given this, the actual Fermi-Dirac distribution of the SNDM particles at the star is given by:
\be
\label{eq:n_start}
dn_*(\vec{p_*})=\frac{1}{e^{\frac{\sqrt{p_*^2+m_X^2}}{T}}+1}
\ee
with $T$ the temperature at the energy sphere. The redshift implies that a minimum momentum is needed in order for the SNDM to overcome the gravitational attraction of the star and escape. This is given by
$p_{*,\text{min}}=\sqrt{\frac{2\Phi m_X^2}{1-2\Phi}}$. Finally, the distribution of the momenta measured at Earth (what was previously called $f(\vec{p})$), can now be written as
\be
f_*(\vec{p})=\frac{1}{e^{\frac{\sqrt{\frac{p^2+m_X^2}{1-2\Phi}}}{T}}+1}
\ee
and the doubly differential rate can expressed, after having transformed the SNDM momentum at the star $p_*$ back to $p$ the SNDM momentum as measured in the lab, as:
\be
\label{eq:ddSNDM}
\frac{dR}{dE~d\cos\theta}=A''  \int\delta(\cos\theta-\frac{q}{2p}) S(q) \sqrt{\frac{p^2+2\Phi m_X^2}{p^2+m_X^2}}f_*(p)~dp
\ee
with $A''$ absorbing constant prefactors.

\subsection{WIMP recoil spectrum}\label{sec:WIMPrecoil}

The full derivation of the doubly differential recoil spectrum for WIMPs can be found in other references~\cite{Gondolo:2002np,LEWIN199687}, so here, we simply state that in the case that we take the Galactic DM halo to be an isothermal sphere, neglect the Earth's motion about the Sun (a reasonable approximation for the purposes of this example), and take into account the Galactic escape velocity ($v_{\text{esc}}$), the recoil spectrum has a parametric dependence of
\begin{widetext}
\be
\label{eq:ddWIMP}
\frac{dR}{dE~d\Omega}= B' S(q)
\left[\exp\left(-\frac{\left(\frac{\sqrt{2 m_A E}}{2 \mu} - v_{\text{lab}}  \cos\gamma\right)^2}{v_p^2}\right)-\exp\left( -\frac{v_{\text{esc}}^2}{v_p^2}\right)\right]
\Theta\left( \cos\gamma - \frac{\frac{\sqrt{2 m_A E}}{2 \mu}-v_{\text{esc}}}{v_{\text{lab}}}\right)
\ee
\end{widetext}
where, as before, $S(q)$ is the structure function of the target nucleus, $m_A$ is the mass of the target nucleus, $E$ is the recoil energy of the nucleus, $\mu$ is the reduced mass of the nucleus and WIMP, $v_{\text{lab}}$ is the speed of the lab in the Galactic frame, taken to be $\sim 232$ km/s, $v_p$ is the peak in the velocity distribution, taken to be 220 km/s, and $\gamma$ is the angle between the recoiling nucleus and $-\vec{v}_{\text{lab}}$ (hence $ d\Omega = d\beta~d\cos\gamma$ for corresponding azimuthal angle $\beta$).

\section{Toy experimental setup}\label{sec:toy}

In this section, we specify a fiducial experimental setup in order to make the conclusions of the previous sections quantitative. Though we will now focus on the detection of a particular example model of SNDM (introduced in Sec.~\ref{sec:toymodel}) in a gas TPC versus WIMP detection in a liquid xenon detector, it must be kept in mind that this is simply an example to illustrate the general advantages of directional sensitivity in discriminating these models.

\subsection{Fiducial experimental parameters}\label{sec:expparameter}

Having defined how we compute our flux in the previous section, we now move to our choice of fiducial detector parameters for this quantitative comparison. Given what has been discussed in the previous sections, in this paper we will consider as benchmark experimental models a liquid-Xe based dual phase TPC and a gaseous TPC with He:CF$_4$ at 1 atm. 

The choice of the former is dictated by the observation that this is the leading approach in WIMP searches above 10 GeV masses and represent the largest existing realizations of a DM detector. It should be noted that similar experiments employing liquid argon, while still currently limited to 50 kg mass~\cite{Agnes:2018fwg}, have demonstrated improved capabilities for electron recoil/nuclear recoil (ER/NR) discrimination with respect to Xe-based detectors and are working on the realization of a 20-ton detector with timelines comparable to the xenon approach~\cite{Aalseth:2017fik}. Nonetheless, our simplified approach to the problem assumes zero ER background and, being based on the experimentally-measured energy profile of the events, can be easily scaled between Xe and Ar by taking into account the differences in the momentum transferred to the nuclei due to the different masses.

Although inherently challenging, gaseous TPCs potentially provide the best architecture and the best observables for directional dark matter detection. Gaseous TPCs can detect the full 3D electron cloud created by an ionization event in the active gas and can measure the total energy of the event through the total collected charge. This implies that they are simultaneously sensitive to the direction and sense (via $dE/dx$) of both electron and nuclear recoils, and that these features can be exploited to discriminate the different kinds of interactions.

We will take He:CF$_4$ as our choice of TPC gas mixture, since at the moment it is the only mixture with simultaneous sensitivity to both spin-independent and spin-dependent couplings that has been demonstrated to have good tracking capability even at 1 atm~\cite{Baracchini:2020nut}. Moreover, we wish to show how very light targets like He are particularly useful not only to explore low WIMP masses, but also in the discrimination between WIMPs and other models like the one discussed in this paper, due to helium atoms' high sensitivity to the transferred momentum. We will assume for our benchmark TPC full 3D tracking capability, including track sense determination. 

It is interesting to note that typical WIMP DM searches have not only a lower energy threshold, but also an upper energy bound on the Region Of Interest (ROI). The reason for this can be discerned from Fig. 1 of Ref.~\cite{Aprile:2018dbl}, where the expected Xe nuclear recoil energy spectra for different WIMP masses are shown together with the experimental detection efficiency and energy ROI selection. As can be seen, the upper ROI limit is chosen to nearly match the maximum possible nuclear recoil energy for a 200 GeV WIMP mass. However, we note that the selection on the energy region can strongly affect the shape of the angular recoil distributions. 

Given that the goal of this paper is to evaluate the capability of DM detectors tailored for WIMP searches to discriminate between WIMPs and models such as the one discussed in the previous sections, we employ in this study an ROI for the Xe-based detector of [4.9, 40,9] keV$_{\text{nr}}$~ \cite{Aprile:2019bbb}. For the gas TPC, since no real underground detectors have been operated yet with such configurations, we extrapolated the lower energy threshold to be 5.9 keV$_{\text{nr}}$ for both He and F recoils from the measurements reported in Ref.~\cite{Baracchini:2020nut}, and the results of simulation discussed in  Ref.~\cite{Vahsen:2020pzb}. For the upper energy thresholds, we used the same assumptions of Ref.~\cite{Aprile:2018dbl} and extrapolated it to be 100 keV$_{\text{nr}}$ for both He and F recoils.

While the chosen lower energy thresholds do not necessarily represent the lowest thresholds achieved by these experimental techniques, we adopt values at which electron recoil discrimination is still significantly effective, since we decide to work under the assumption of zero background.
 
Similarly, we extracted the experimental energy resolution from measurements on actual data. In particular, we assumed the Xe-based detector energy resolution dependence follows the relation $\sigma_{E}(E) = a/\sqrt{E} + b$ with $a$ and $b$ taken from Table III of Ref.~\cite{Aprile:2017xxh}. For the gas TPC, we adopted the function shown in Ref.~\cite{Vahsen:2014fba} describing the relative gain (and therefore energy) resolution as $\sigma_{G}(G) = \sqrt{d^2 + c^2/E}$ with d = (1.94$\pm$0.07), and c = (22.3$\pm$1.5) $\sqrt{\rm{keV}}$, with the constraint to reproduce the 18$\%$ energy resolution at 5.9 keV$_{\text{ee}}$ reported by the He:CF$_4$ detector in Ref.~\cite{Costa:2019tnu} and 2$\%$ above 50 keV$_{\text{ee}}$ as in typical gas detectors~\cite{Vahsen:2014fba}. 

Gaseous TPC angular resolution is constrained at very low energies (below about 50 keV$_{\text{nr}}$) mainly by multiple scattering, straggling, and diffusion during drift. Given that no measurements exist of angular resolutions from 3D DM TPCs in a realistic regime (in terms of underground operation of detectors of $\mathcal{O}$(1) m$^3$ dimensions), and since we want to be as general as possible, we perform the analysis with a wide range of possible resolutions, spanning from 2$^{\circ}$ to 45$^{\circ}$. The former is an almost perfect angular resolution which represents the ideal case when all angular information is available to the experiment, neglecting the aforementioned diffusion, straggling, and multiple scattering. The latter reflects a scenario of low resolution where a hemisphere in Galactic coordinates is split into a handful of distinct bins. This last assumption is backed up by measurements in the 50-400 keV$_{\text{nr}}$ range by the NEWAGE experiment \cite{Nakamura:2012zza} and is consistent with the simulation studies presented in Ref.~\cite{Vahsen:2020pzb}.

Furthermore, we will take our fiducial experimental scenarios to have perfect background rejection. Our interest is \textit{not} in the detection of a signal, but in the subsequent discrimination between two models after a discovery. In order not to obfuscate this point, we henceforth assume the capability of our fiducial experimental setups to fully reject any other background sources through typical analysis techniques. As a corollary of this, we will also not be interested in the comparative exposure of the experiments and the specific cross-sections of the different models, instead selecting cross-sections and exposures that make discrimination maximally difficult.

We would nonetheless like to stress that, while the details of the background are different for every experimental setup, the energy spectrum of backgrounds in direct detection experiments often highly resembles the spectrum expected from the signal, while the angular distribution does not due to the general isotropy of the background sources or clear directional point source (e.g. the Sun).
This provides an important means of background rejection in directional detectors that we have neglected here, a consideration of which would only serve to improve the relative discrimination capabilities between DM models of a directional detector over a non-directional detector. A full treatment of DM detection and discrimination in the presence of backgrounds will be the subject of an upcoming paper.

\subsection{Signal scenarios for comparison}

\begin{table*}
\begin{center}
\begin{tabular}{c|c|c|c|c|c|c}
Scenario & Target & WIMP Mass [GeV]  & SNDM Mass [MeV]  & $T$ [MeV] & $\log_{10} y$  & $\Phi$   \\ 
\hline
\hline
1 & $^4$He & 10  & 5& 0.31 & -13.3 & 0.006 \\ 
2 &$^{19}$F & 10 & 7 & 1.0 & -14.3 & 0.02 \\ 
3 &$^{131}$Xe & 10 & 9 & 1.6 & -14.6 & 0.03 \\ 
4 &$^4$He & 100 & 5 & 0.52 & -14.0 & 0.01 \\ 
5 &$^{19}$F & 100 & 14 & 3.0  & -15.0 & 0.07 \\ 
6 &$^{131}$Xe & 100 & 38 & 13.4 & -16.0 & 0.1
\end{tabular}
\caption{Various signal scenarios for comparison. We present here the target nucleus, WIMP mass, SNDM mass, SNDM-SM coupling $y$, temperature of SNDM particles at the energy sphere set by this $y$, and the redshift factor $\Phi$ from this energy sphere. The WIMP masses and SNDM parameters are chosen in order to have comparable momentum transfer to the nuclei such that energetic discrimination is difficult, as discussed in Sec.~\ref{sec:kinematic}.}
\label{tab:sn_wimp_cases}
\end{center}
\end{table*}

In this section, we present the various scenarios we consider as test cases for discrimination between a WIMP and SNDM signal in our fiducial experimental setups.

Since the argument we make in this paper is that a WIMP signal and SNDM signal are very difficult to distinguish through solely the nuclear recoil energy spectra, we considered pairs of WIMP masses and SNDM scenarios (mass, SNDM-SM coupling $y$, corresponding temperature at escape $T$, and associated redshift factor $\Phi$) that will produce a similar energy deposition in a DM detector. 
 
As shown in Sec.~\ref{sec:kinematic}, order to obtain this, the WIMP $\mu v_0$ needs to match the SNDM $p_0$ such that the WIMP and SNDM have comparable momentum transfer to the target nucleus. We have chosen an example of lighter (10 GeV) and heavier (100 GeV) WIMP mass to compare to various SNDM scenarios. Since the energy and angular recoil distribution are proportional to the WIMP-nucleus cross-section, we allow this to scale freely to best match the energy spectrum produced by the SNDM. As per the fiducial experimental choices discussed in Sec. \ref{sec:toy}, we select $^4$He, $^{19}$F, and $^{131}$Xe as target elements and consider six different scenarios for comparison, where a ``scenario'' is a choice of WIMP mass, SNDM mass and coupling, and target nucleus. These scenarios are listed in Table~\ref{tab:sn_wimp_cases}. We believe that our choices are a good representation of the various cases that may be encountered in terms of experimental approaches, target materials, and DM parameter space.

\begin{figure*}[!th]
  \centering
\includegraphics[width=0.45\textwidth]{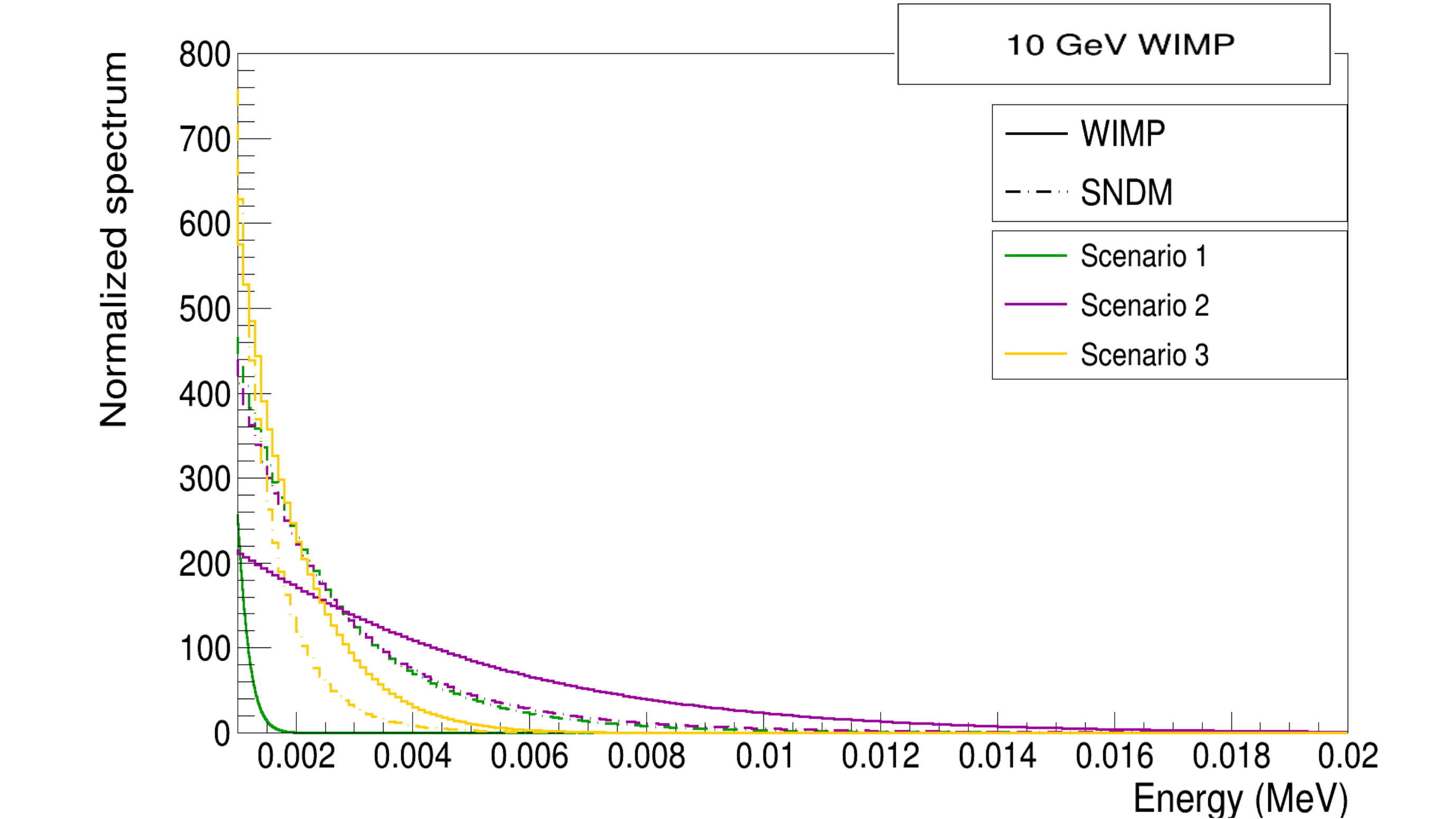}
\includegraphics[width=0.45\textwidth]{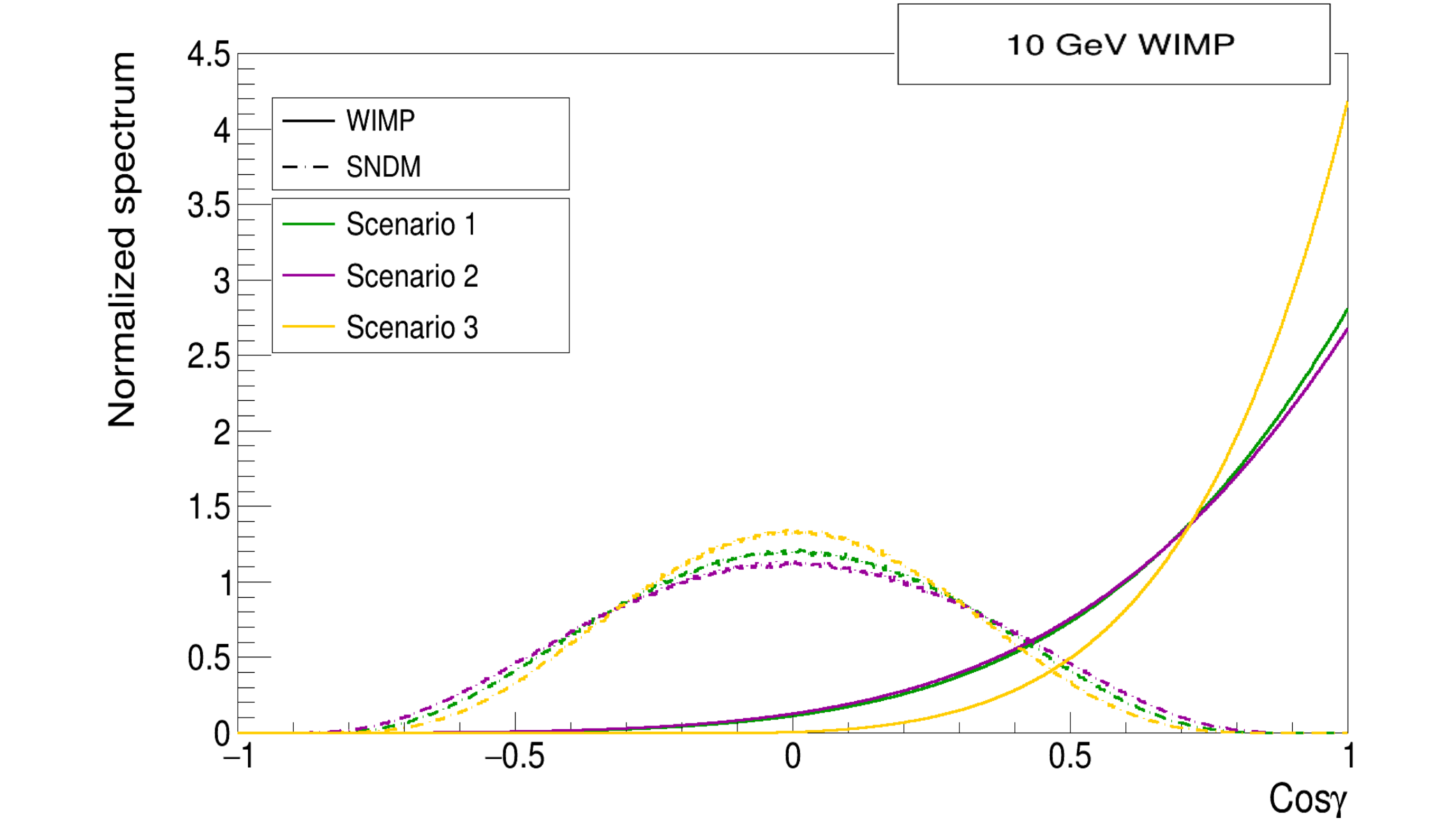}
\includegraphics[width=0.45\textwidth]{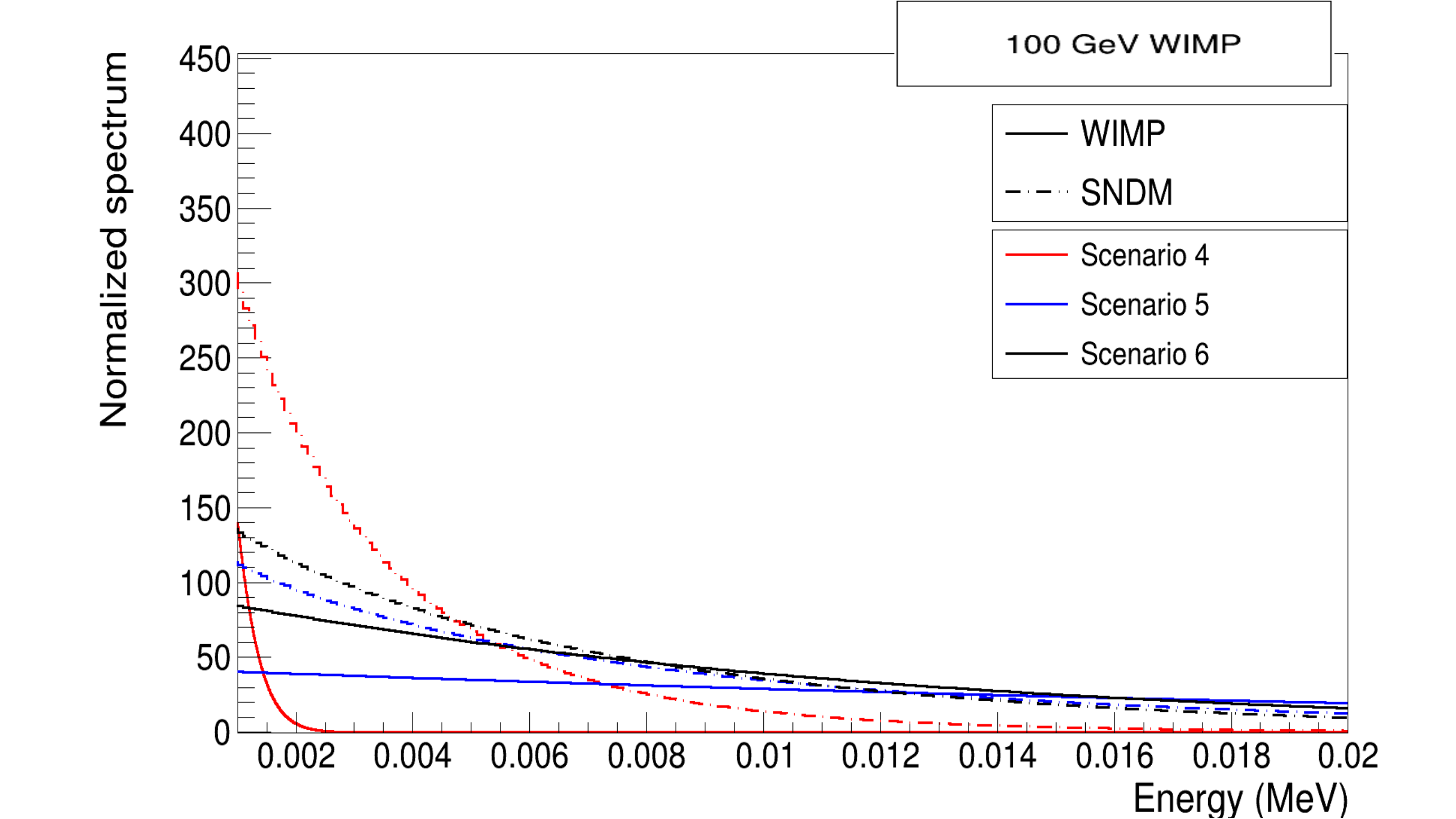}
\includegraphics[width=0.45\textwidth]{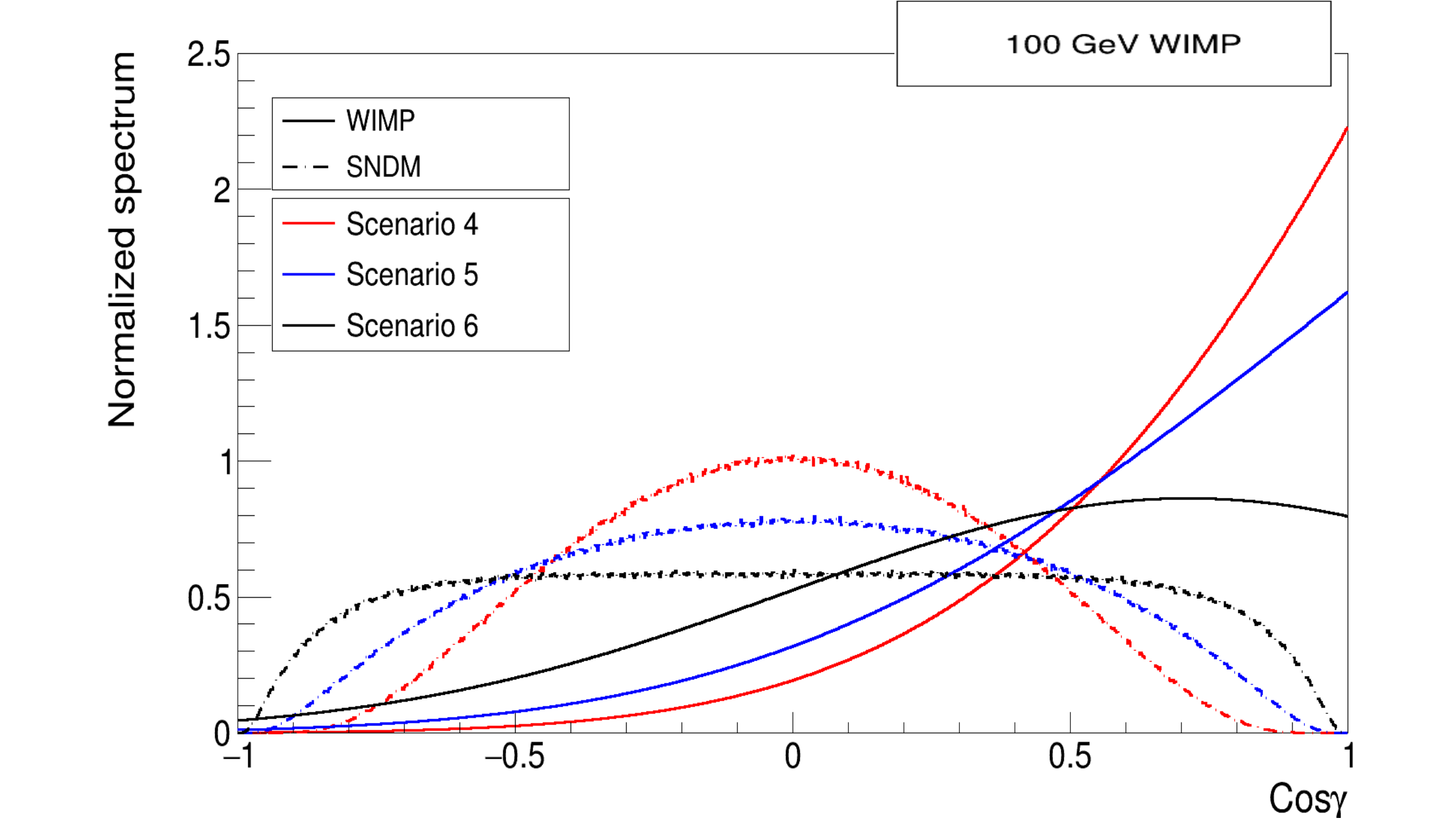}
 \caption{Energy (left) and 1D angular recoil spectra (right) for 10 GeV (top) and 100 GeV (bottom) WIMP masses for the six scenarios considered in this paper. (See Table~\ref{tab:sn_wimp_cases}.) Solid lines denote the WIMP spectrum while dashed lines denote the SNDM spectrum; color denotes the scenario in question. Note the similarity in energetic spectral shape for WIMPs and SNDM of the same scenario (owing to our choice of WIMP and SNDM masses to deposit similar energy in nuclear recoils) and the dramatic difference in angular spectrum by virtue of the roughly perpendicular arrival directions.}
 \label{fig:spectra}
 \end{figure*}
 
 Fig.~\ref{fig:spectra} shows the energy (left) and 1D angular (right) distributions of nuclear recoils for the six scenarios considered, divided into 10 GeV WIMP scenarios (top) and 100 GeV scenarios (bottom). These spectra are shown after having applied the cut on the energy ROI, as discussed in Sec.~\ref{sec:expparameter}. Here $\gamma$ is the angle between the laboratory velocity and recoiling nucleus in the lab frame, as from Eq.~\ref{eq:ddWIMP}. The full 2D angular spectra are shown in Appendix~\ref{app:2d}.
 
 As is evident in the plots, the energy spectra possess very similar shapes for a given scenario, as expected by our choice of WIMP mass and SNDM parameters. The angular distributions, on the contrary,  demonstrate a dramatic difference in shape due to the approximately perpendicular arrival directions of the WIMP and SNDM. It is this difference that allows the angular spectra to discriminate between the two models with very few events.

\section{Results}
\label{sec:results}

\begin{figure*}[t]
  \centering
 \includegraphics[width=0.47\textwidth]{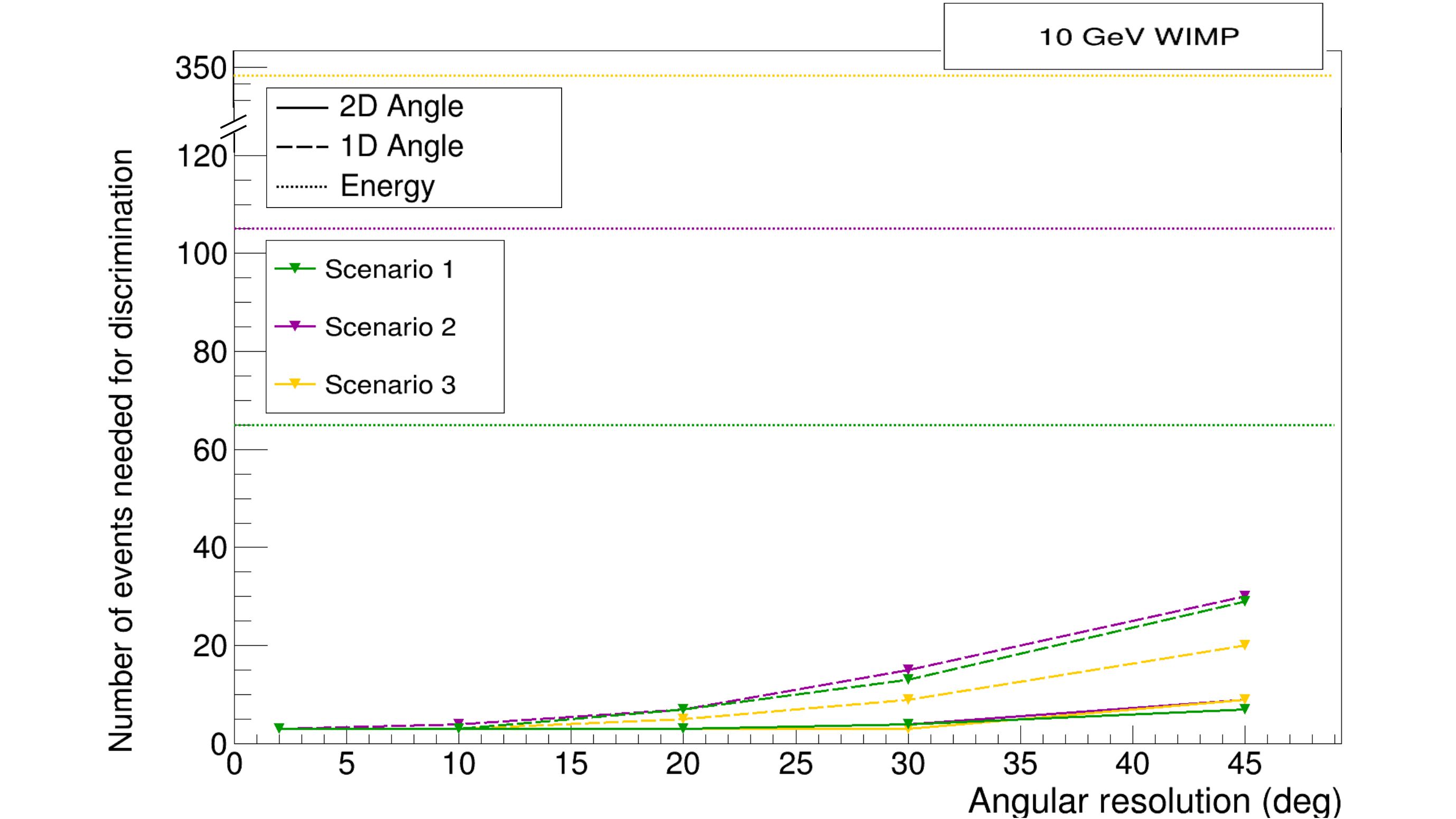}
 \includegraphics[width=0.48\textwidth]{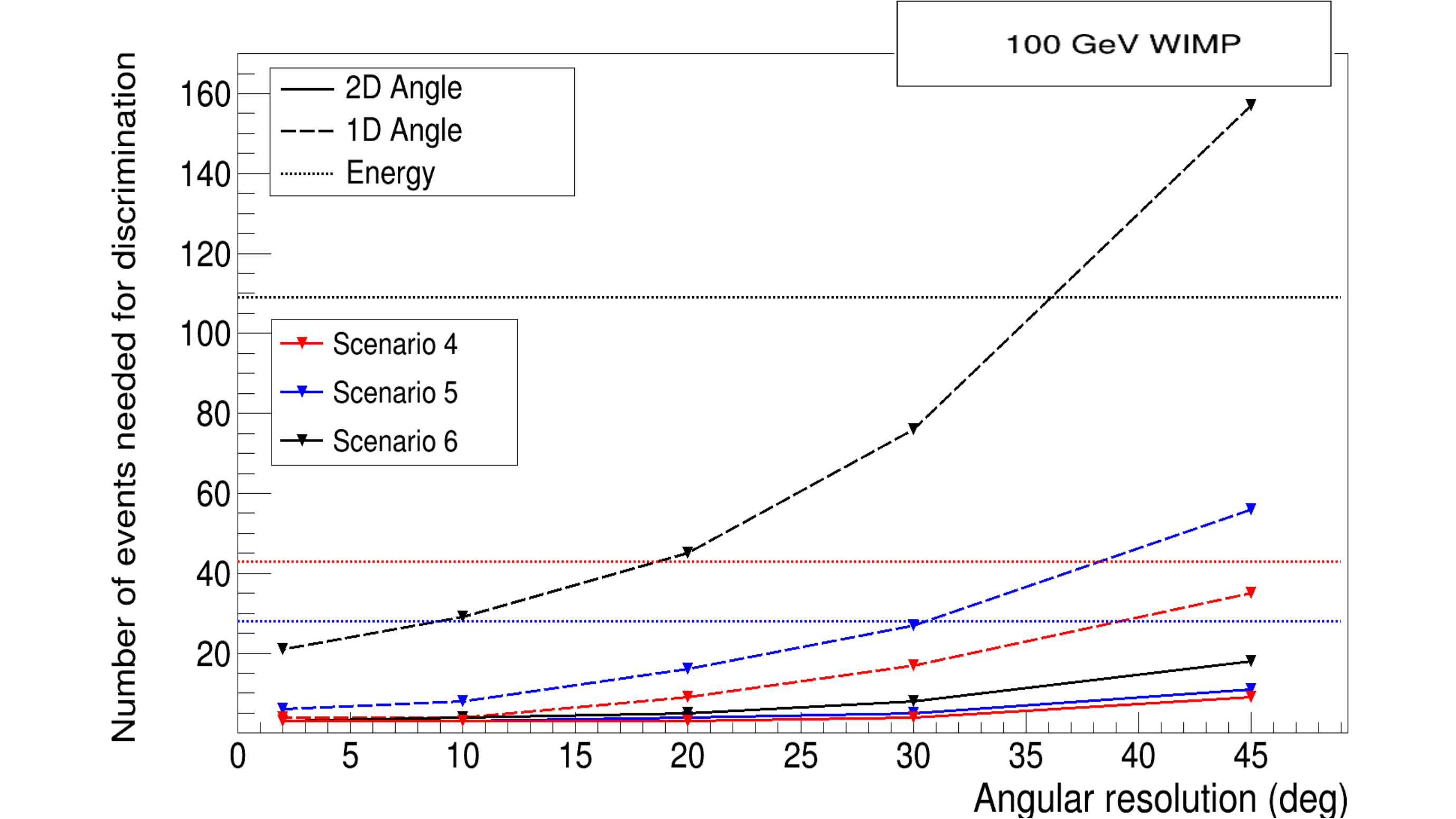}
 \caption{The average number of events necessary for discriminating between a WIMP and SNDM signal in our fiducial experimental setups for the various Scenarios in Table~\ref{tab:sn_wimp_cases}, plotted as a function of angular resolution. (n.b. For the 2D spectrum, the horizontal axis has units of deg$~\times~$deg for each point, e.g. ``30'' denotes $30^{\circ} \times 30^{\circ}$.) Results for energy spectra are horizontal as they possess no dependence on angular resolution. It is clear that angular information allows model discrimination with far fewer average signal events than when using solely the energy spectra. While the specific numeric values are dependent on our choice of fiducial experimental setup, the point remains that directional detectors can provide an order of magnitude improvement in discrimination capability for a large selection of realistic angular resolutions.}
 \label{fig:relall}
 \end{figure*}
 
In order to estimate the improvement in signal discrimination using angular information, we compute the approximate number of events to discriminate a WIMP and SNDM spectrum for our various scenarios (Table~\ref{tab:sn_wimp_cases}) in our fiducial experimental setups (Sec.~\ref{sec:toy}). This gives a good demonstration of the power of the angular spectrum in discriminating between these models in comparison to purely energetic spectra.

The statistical technique chosen to compute the discrimination power is based on a profile likelihood ratio test in which we find the average number of detected events with either energy, 1D angle, or 2D angle spectral information needed to discriminate between the two different models with $<5\%$ chance of committing a Type I or Type II error. For the full details of the Monte Carlo data generation and subsequent statistical analysis, see Appendix~\ref{app:stat}.
 
The number of detected events inside the ROI needed to distinguish between the WIMP and SNDM models are shown in Figure \ref{fig:relall} for the six scenarios considered as a function of angular resolution. We show the average number of events to distinguish the models using purely the energy spectra as a dotted line (horizontal, as there is no angular resolution dependence), the number events for using only the 1D angular spectrum as a dashed line, and the number of events using only 2D angular spectrum as a solid line. The Tables \ref{tab:results2}-\ref{tab:results45} in Appendix~\ref{app:tables} tabulate the same results numerically, though the actual values are less important than the clear order of magnitude difference in required events using the energy and angular spectra.

Even in the worst case scenario with 45$^{\circ}$ angular resolution, for which only a handful of bins are available for the determination of the angular distribution of the detected events, the angular spectra are still far more powerful at discriminating the two hypotheses than the energy spectra. As expected, a full 2D angular detection performs much better than its projection on a single axis, as the 2D provides more information on the original shape of the recoil distribution. This allows model discrimination with more than one order of magnitude fewer events compared to using the energy spectra, especially when a good resolution is available or if the kinematics are particularly favorable, as for the Scenario 3, where the SNDM and WIMP distributions are strongly peaked in different directions.

As expected, the angular resolution significantly affects the discriminating power. Indeed, a worse angular resolution allows the events to migrate away from their expected position, effectively diluting and smearing the original shape of the recoil distributions. For the 1D angular spectra, the number of events needed to distinguish seems to double roughly every 10$^{\circ}$ lost in resolution, resulting in a strong degradation of the information from a resolution of 30$^{\circ}$ or worse. For the 2D angle, worsening the resolution has a lesser effect, with a clear degradation of the discrimination power noticeable only at resolutions worse than $30^{\circ}\times30^{\circ}$. Moreover, the relative increase in number of events is less intense than the 1D case.

Another interesting feature appearing in the right-hand plot of Figure~\ref{fig:relall} is that energy distributions tuned to be similar in terms of total momentum transfer can still exhibit differences in shape that allow better discrimination than with 1D angular information alone, as for the case of $^{19}$F target with a WIMP mass of 100 GeV (Scenario 5). A similar effect is happening also for the $^{131}$Xe target with 100 GeV WIMP mass (Scenario 6), but in this case this is due to the well-known ringing effect~\cite{Bozorgnia_2012} that results in more similar angular distributions. Nonetheless, it is important to notice that this happens only in the case of very poor angular resolution, and is not present when this is improved beyond 30$^{\circ}$. This underlines the fact that such a feature is only an artifact of our simplistic analysis in which we do not use the angular and energy distributions together, with the proper correlations induced by the kinematics, as a real experiment would do. 

Overall, the results show that a directional detector would possess a strong advantage in distinguishing between a WIMP and SNDM model even with very few signal events.

\section{Conclusions}
\label{sec:conc}

In this paper, we have shown that directional dark matter detection would provide a crucial ability to discriminate between a cold, cosmological population of GeV-scale WIMP dark matter and a hot, supernova-produced population of MeV-scale dark matter. This is yet another, heretofore unrecognized, motivation to continue research and development into directional dark matter detection. If we wish to not only detect dark matter, but determine its origin and properties, directional detection will be the defining tool to do so. \\

\begin{acknowledgments}
WD would like to thank his collaborators Peter W. Graham, Dan Kasen, Gustavo Marques-Tavares, and Surjeet Rajendran for their contributions to the initial result on producing a diffuse Galactic flux of DM with supernovae.

WD is supported in part by DOE Grant DE-SC0012012, by NSF Grant PHY-1720397, the Heising-Simons Foundation Grants 2015-037 and 2018-0765, DOE HEP QuantISED award \#100495, and the Gordon and Betty Moore Foundation Grant GBMF7946.

EB and GD are supported in part by the European Research Council (ERC) under the European Union’s Horizon 2020 programme (grant agreement No 818744).
\end{acknowledgments}

\begin{appendix}
\section{Alternative discrimination strategies}
\label{app:elec}

While we have focused mainly on nuclear recoil experiments and directional detection with gas TPCs, one may also wonder if there are not other ways to discriminate the WIMP and SNDM signals. For example, one may wish to discriminate SNDM from WIMPs with electron recoils. In this case, the resulting recoil spectra are significantly different for the two models. In the limit where $m_X \gg m_A$, Eq.~\ref{eq:momentum} becomes $|\vec{k}_{\text{elec}}| \approx 2m_A v_0 \cos\theta_r$. The critical difference here is that there is now only a dependence on the incident \textit{velocity}, not the momentum. Hence a recoil due to a cold WIMP at $v_0 \approx 10^{-3}c$ would yield only a $m_A v_0^2 \sim 1$ eV recoil energy while hot SN-produced DM at $v_0 \sim c$ would yield recoil energy $2 m_A v_0$ up to $\sim 1$ MeV, six orders of magnitude higher.

However, this approach suffers from many drawbacks. Detection via electron scattering is made difficult by a low cross-section in comparison to nuclei. Generically, the cross-section will be lower due to the lack of the coherent enhancement factor of the nucleus ($Z^2$) which is $\mathcal{O}(10^3)$ for nuclei of interest and by a kinematic suppression for scattering off the electrons.
Additionally, electron recoil detectors have no ability to discriminate SNDM from, e.g., some additional hot isotropic population of new particles.

Another possible means of discrimination would be via the annual modulation signal, since in the case of the SNDM, the semi-relativistic velocities wash out any dependence on the relative motion of the Earth. Hence, one would expect to see no modulation whatsoever. Once again, though, this would not allow discrimination from some hot cosmological abundance of new particles. A time-independent signal is also much harder to isolate from background.

Finally, one might hope to discriminate between an SNDM and WIMP signal by comparing the energy spectra in various target media (as in Ref.~\cite{Billard:2013qya}, in which the authors suggest this technique in order to discriminate a WIMP signal from neutrino background), however this would require experiments with very large exposures to generate sufficient signal statistics.

As a result of the above considerations, we chose to focus on nuclear recoils in truly directional detectors in this study.

\section{Two-dimensional signal plots}
\label{app:2d}

\begin{figure*}[!th]
  \centering
\includegraphics[width=0.35\textwidth]{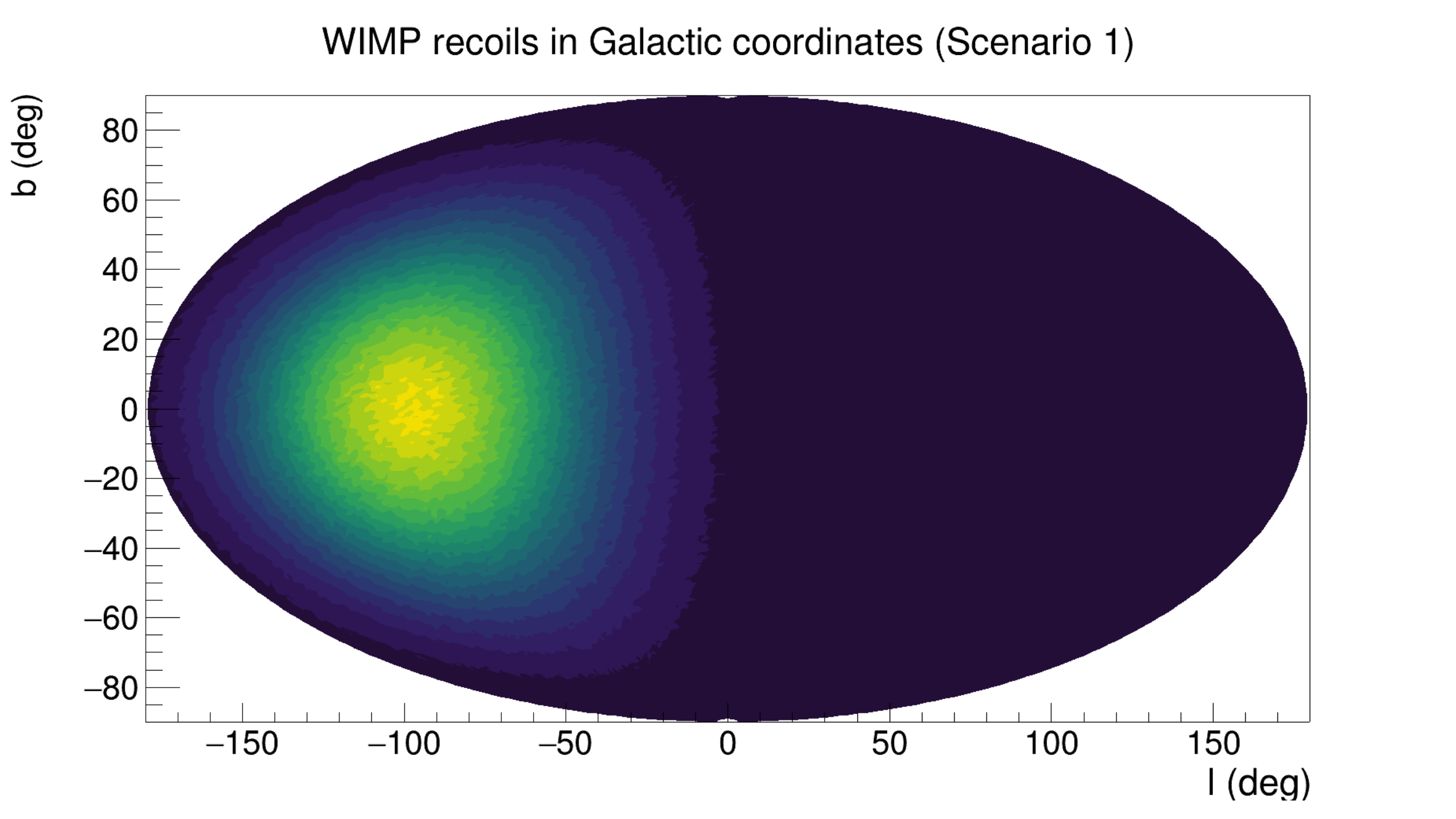}
\includegraphics[width=0.35\textwidth]{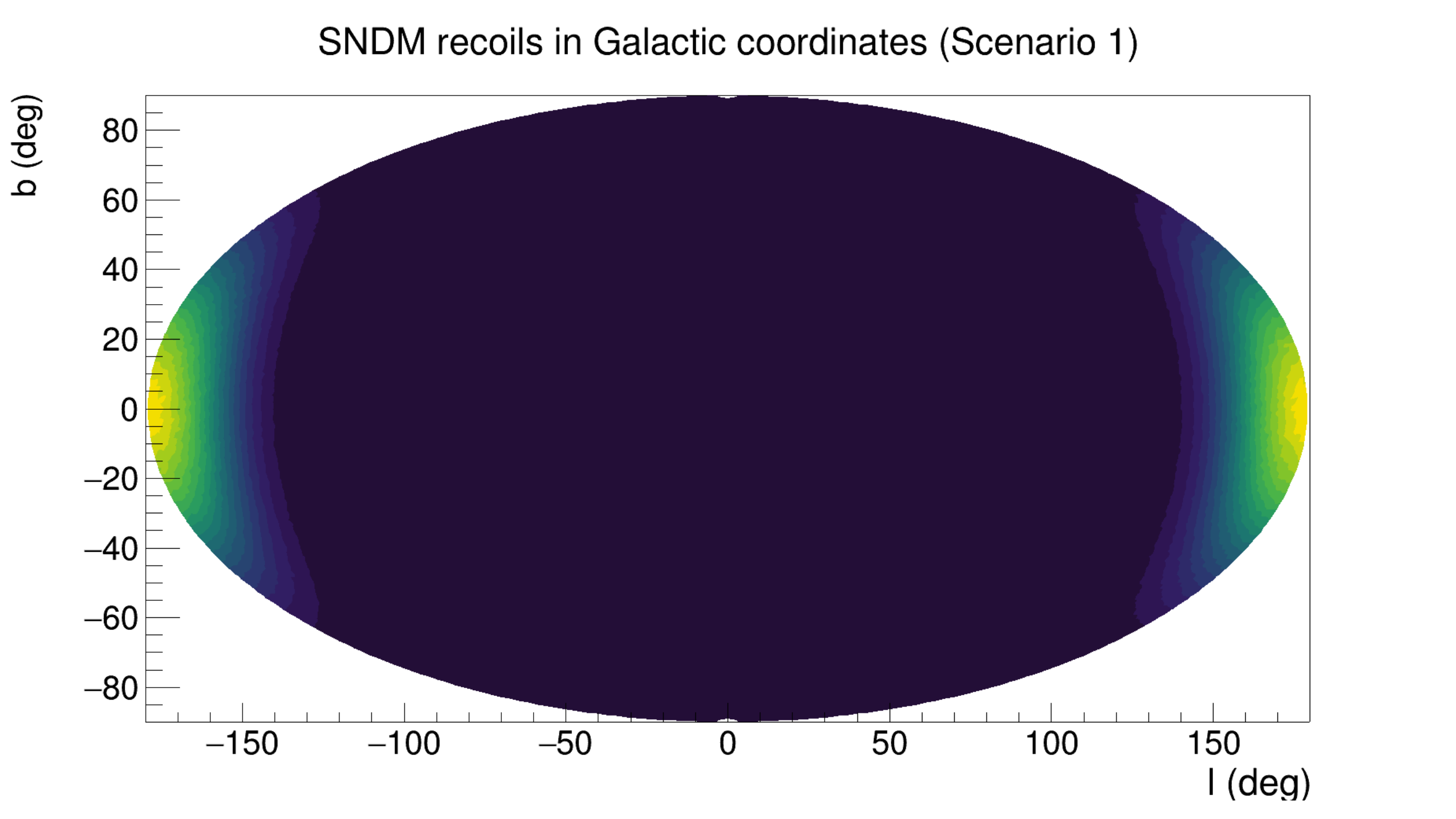}
\\
\includegraphics[width=0.35\textwidth]{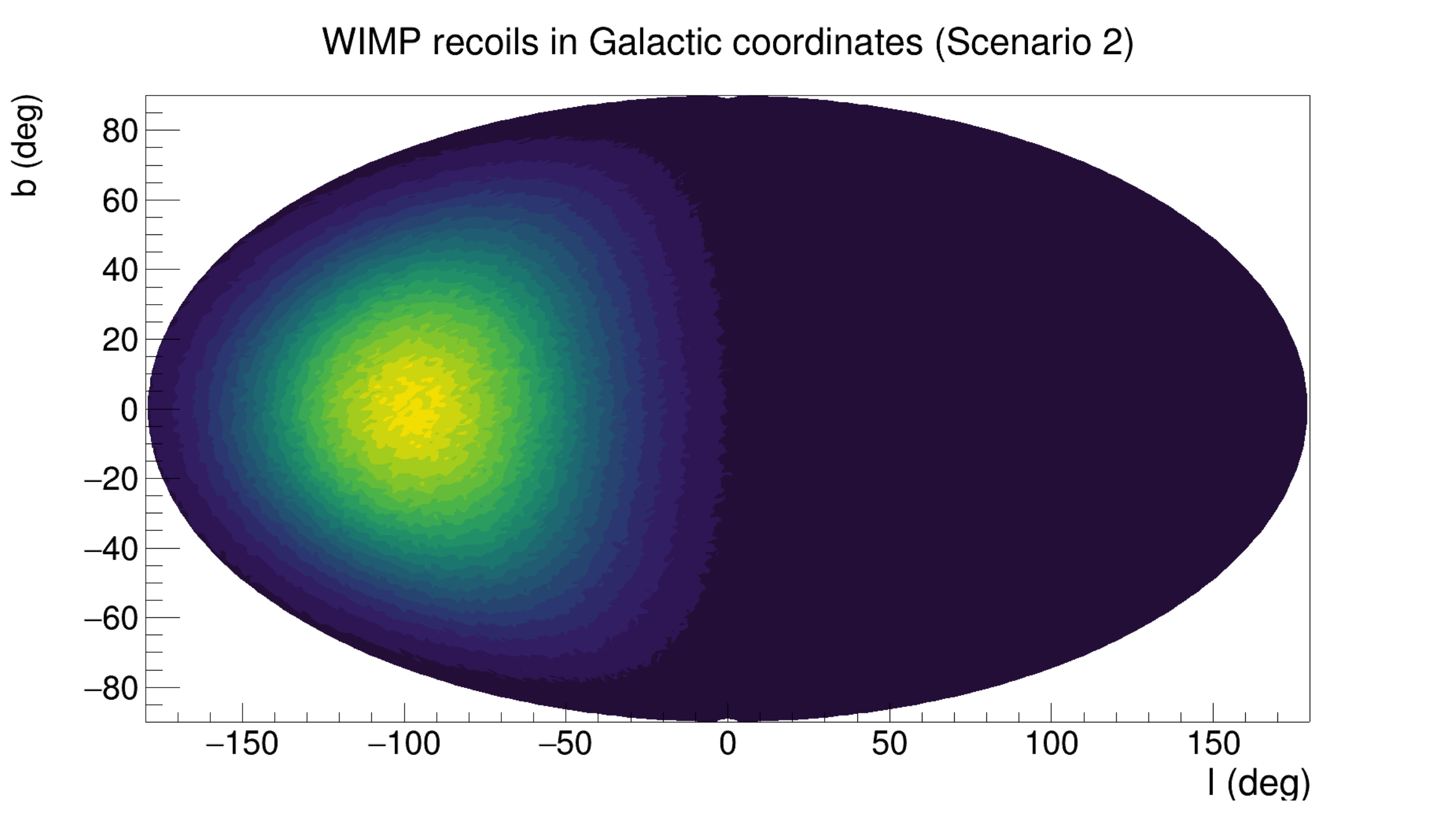}
\includegraphics[width=0.35\textwidth]{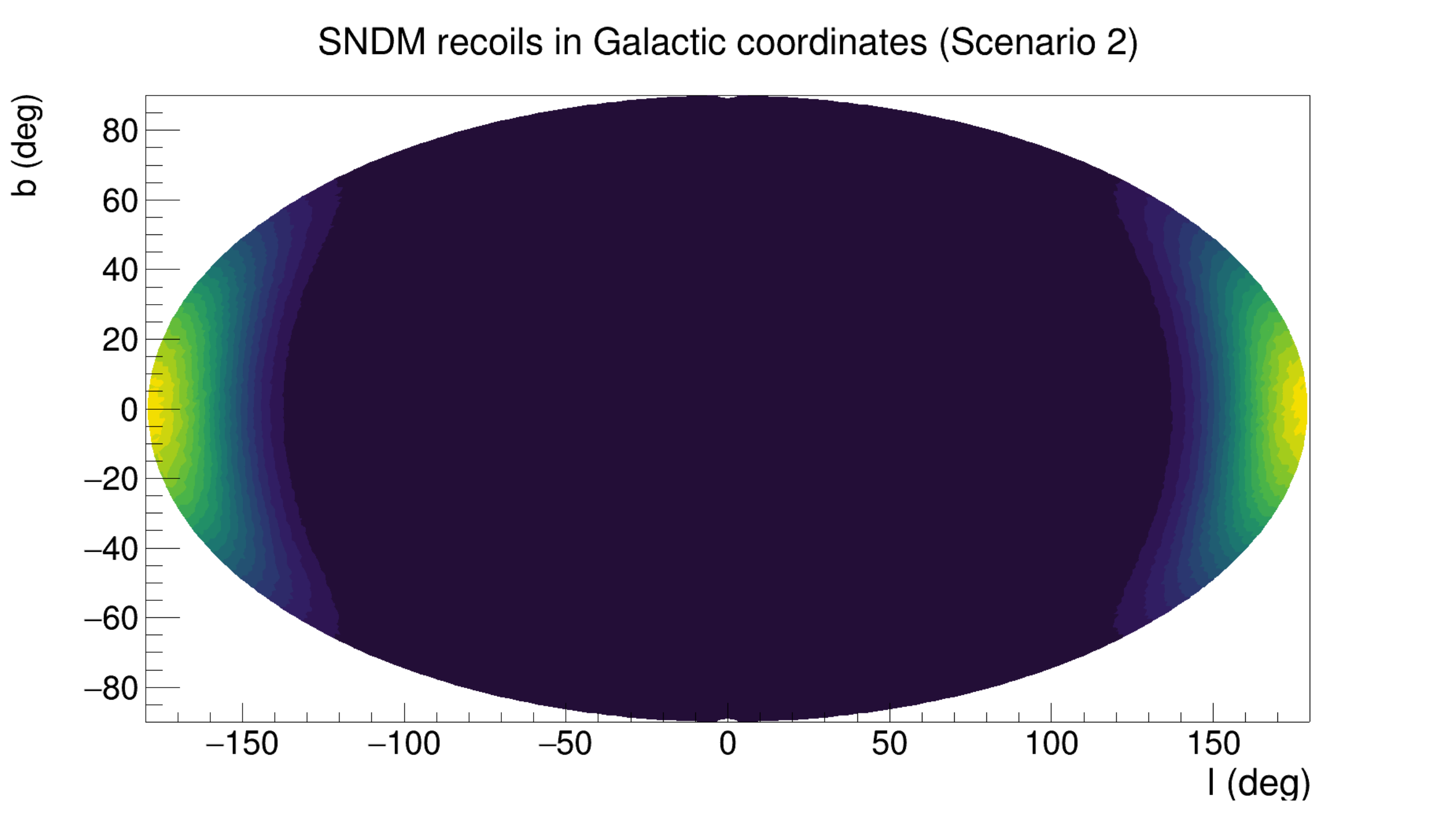}
\\
\includegraphics[width=0.35\textwidth]{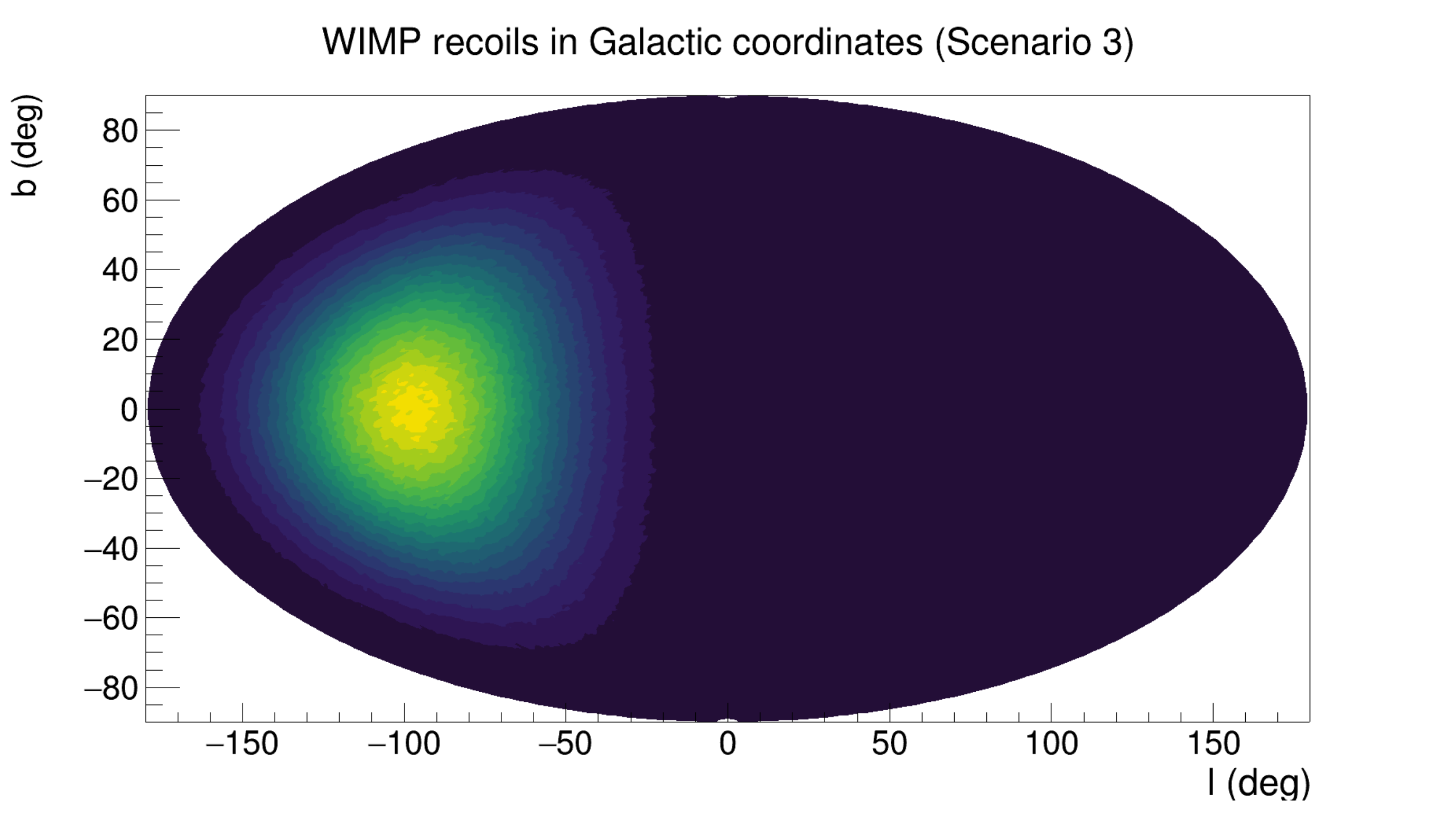}
\includegraphics[width=0.35\textwidth]{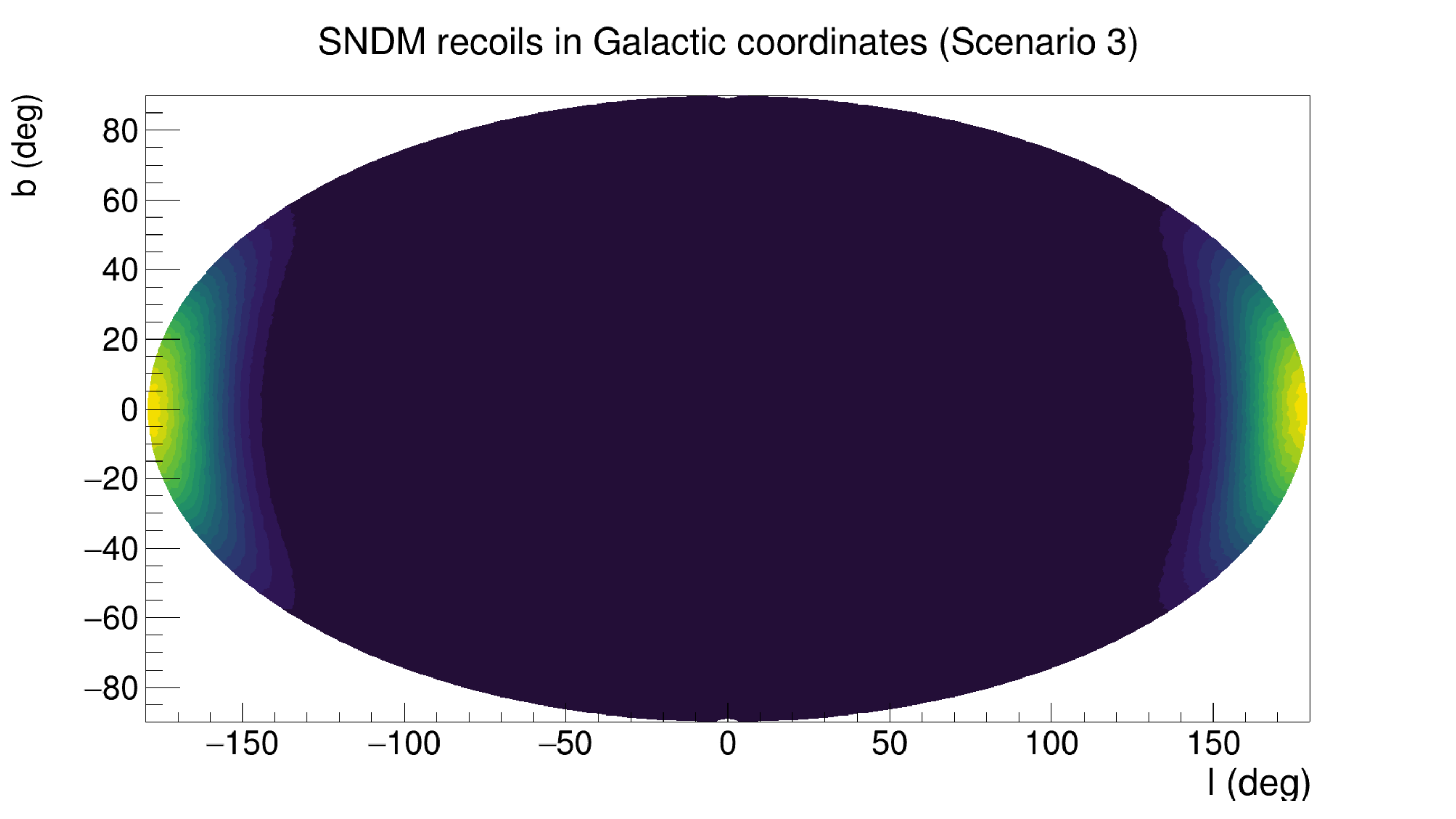}
\\
\includegraphics[width=0.35\textwidth]{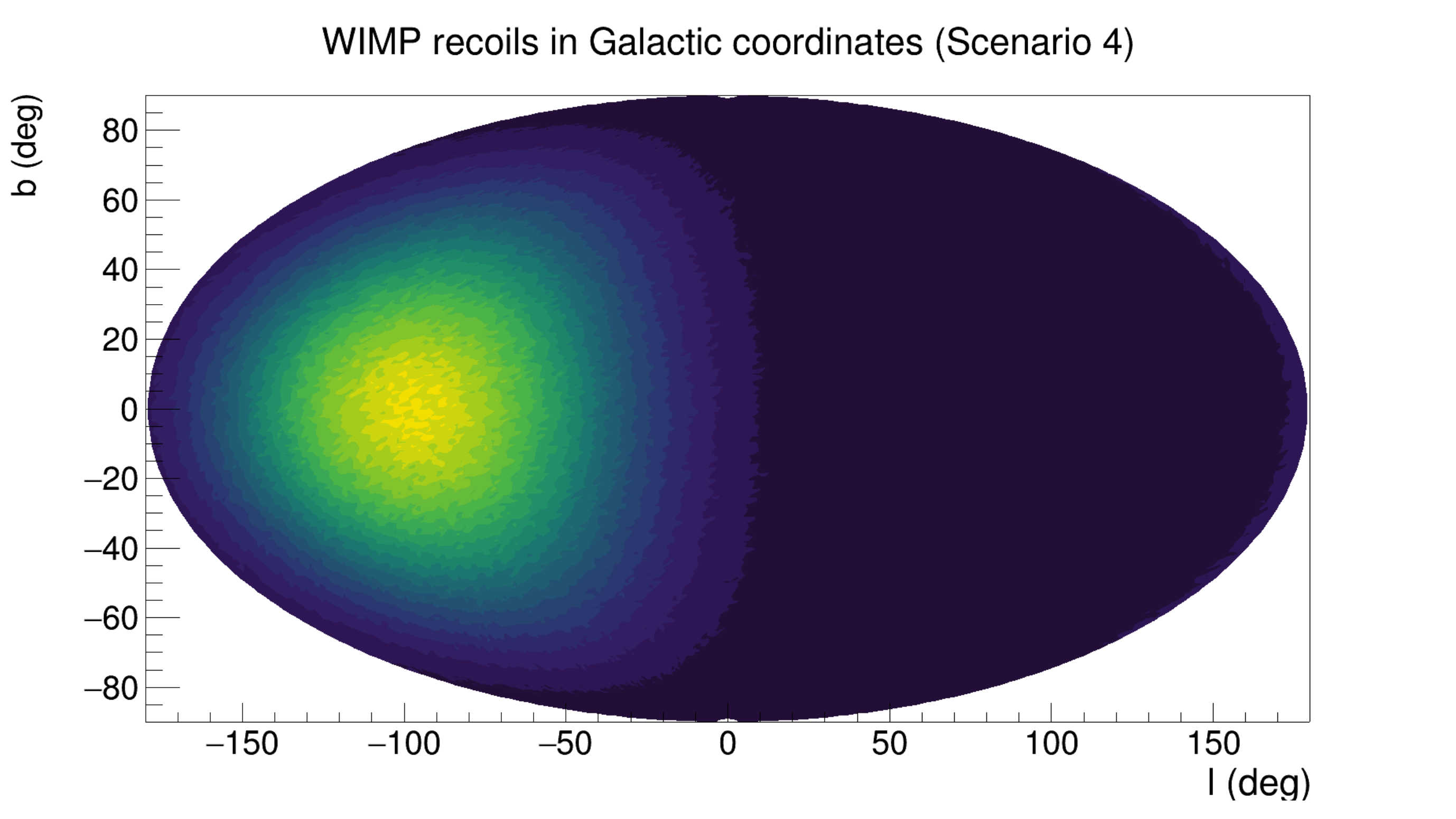}
\includegraphics[width=0.35\textwidth]{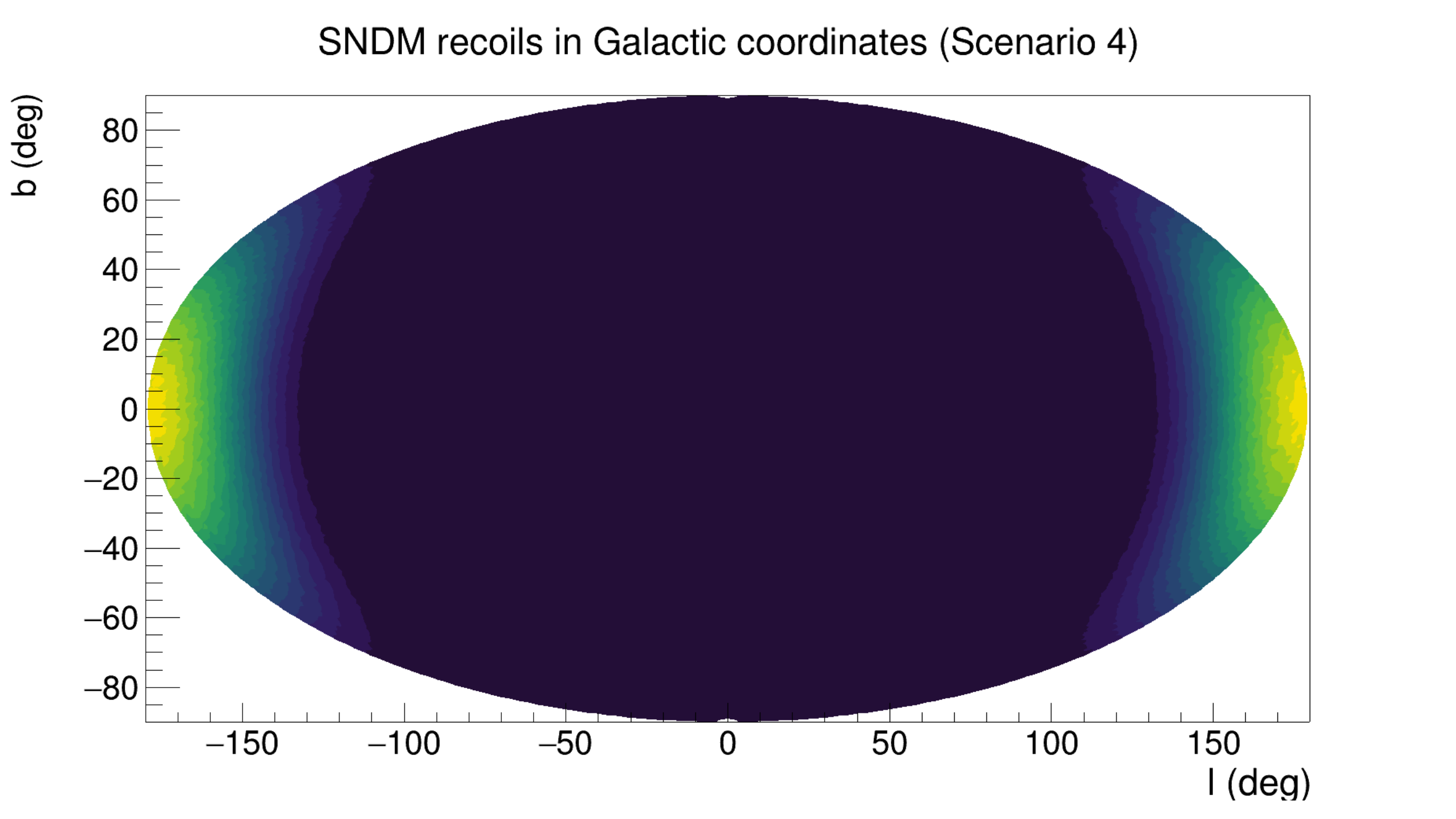}
\\
\includegraphics[width=0.35\textwidth]{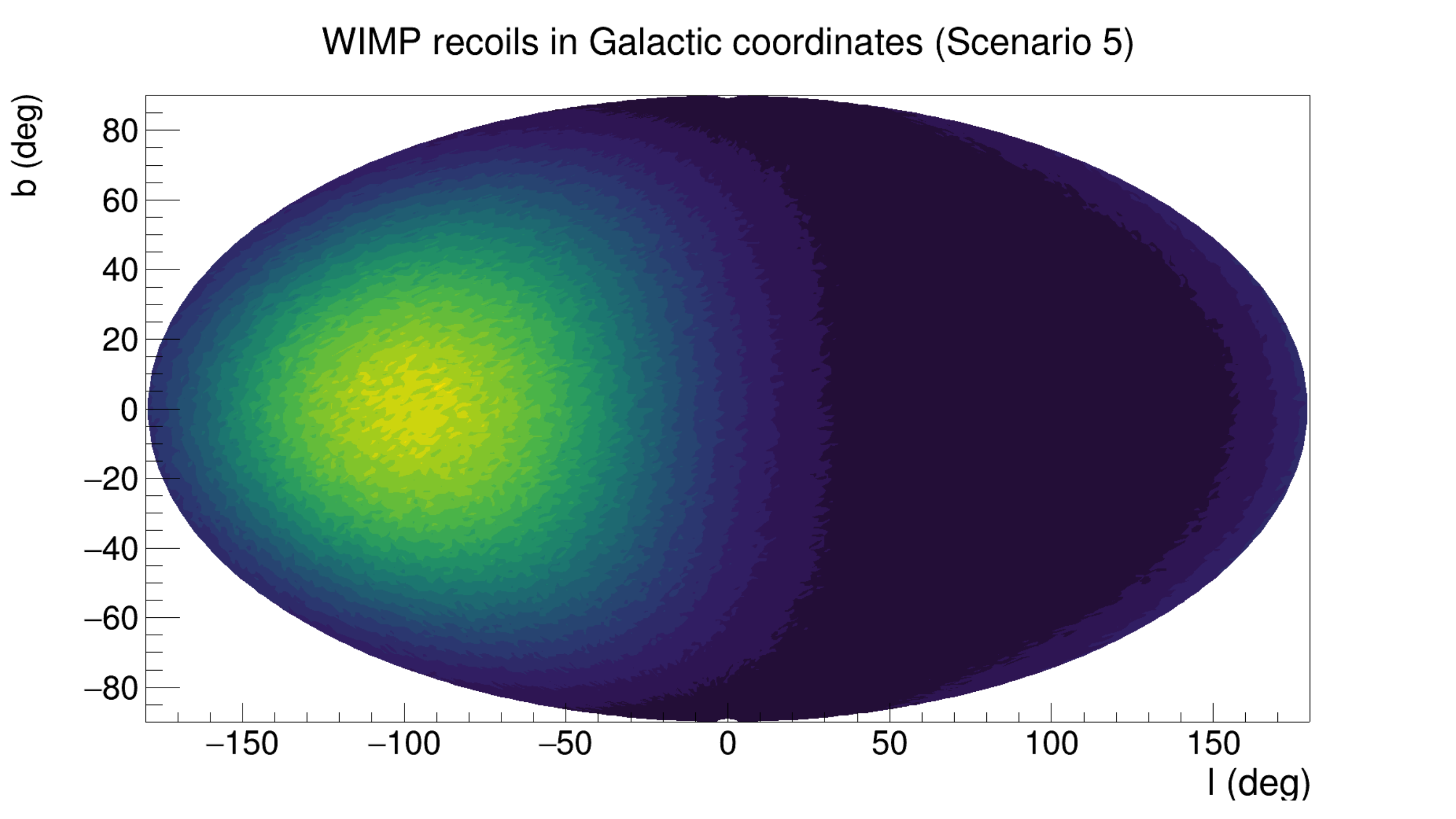}
\includegraphics[width=0.35\textwidth]{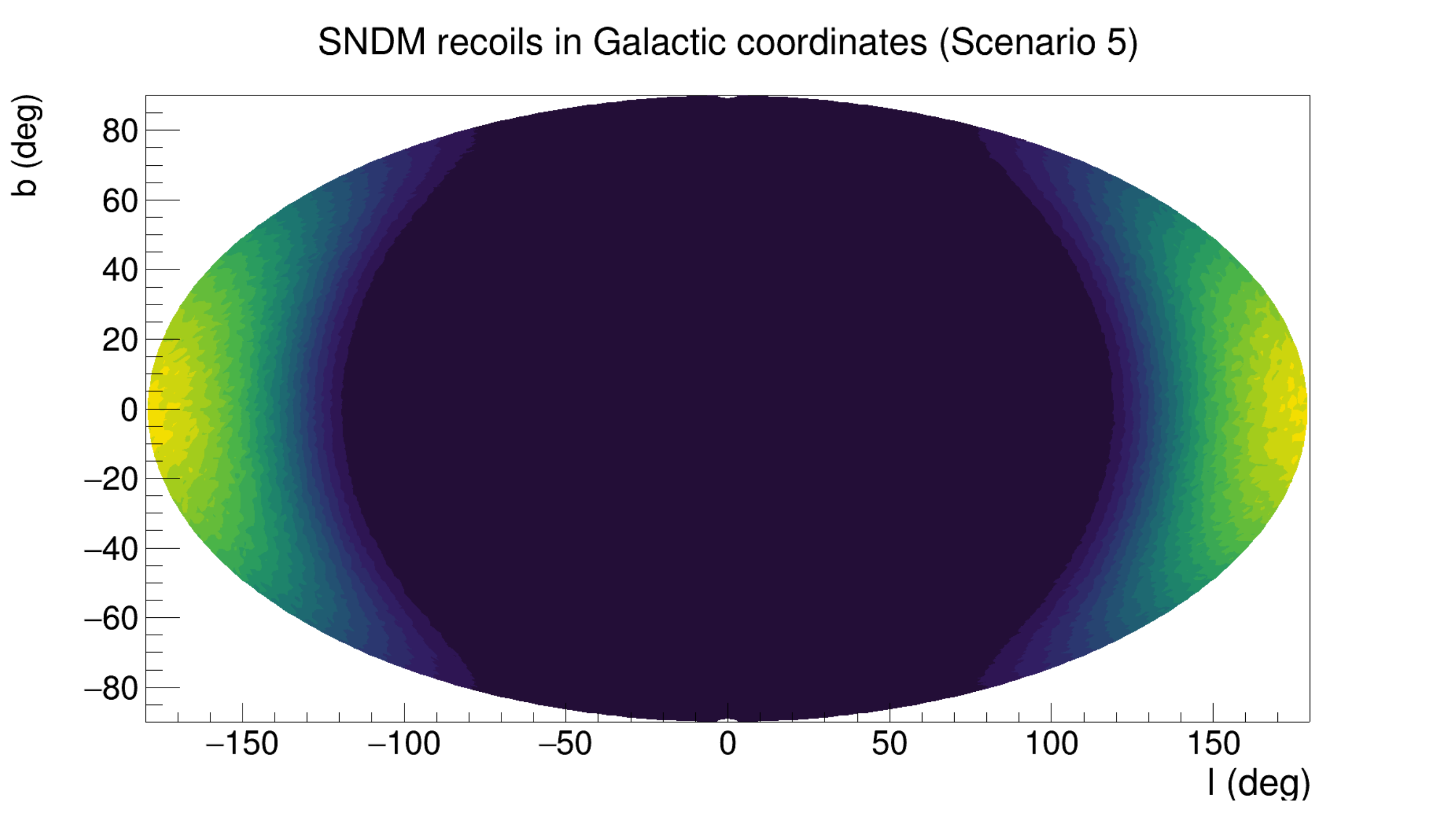}
\\
\includegraphics[width=0.35\textwidth]{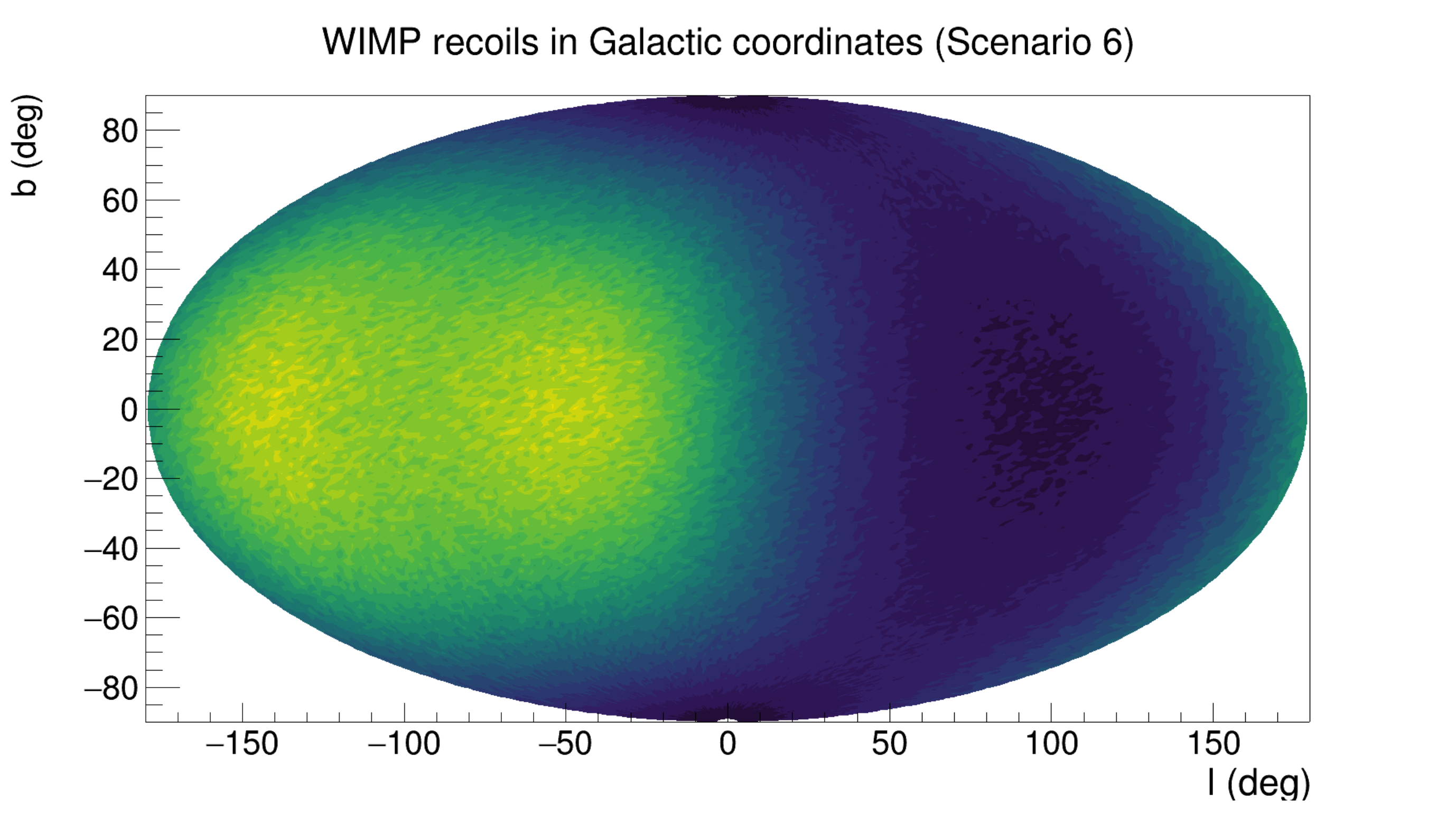}
\includegraphics[width=0.35\textwidth]{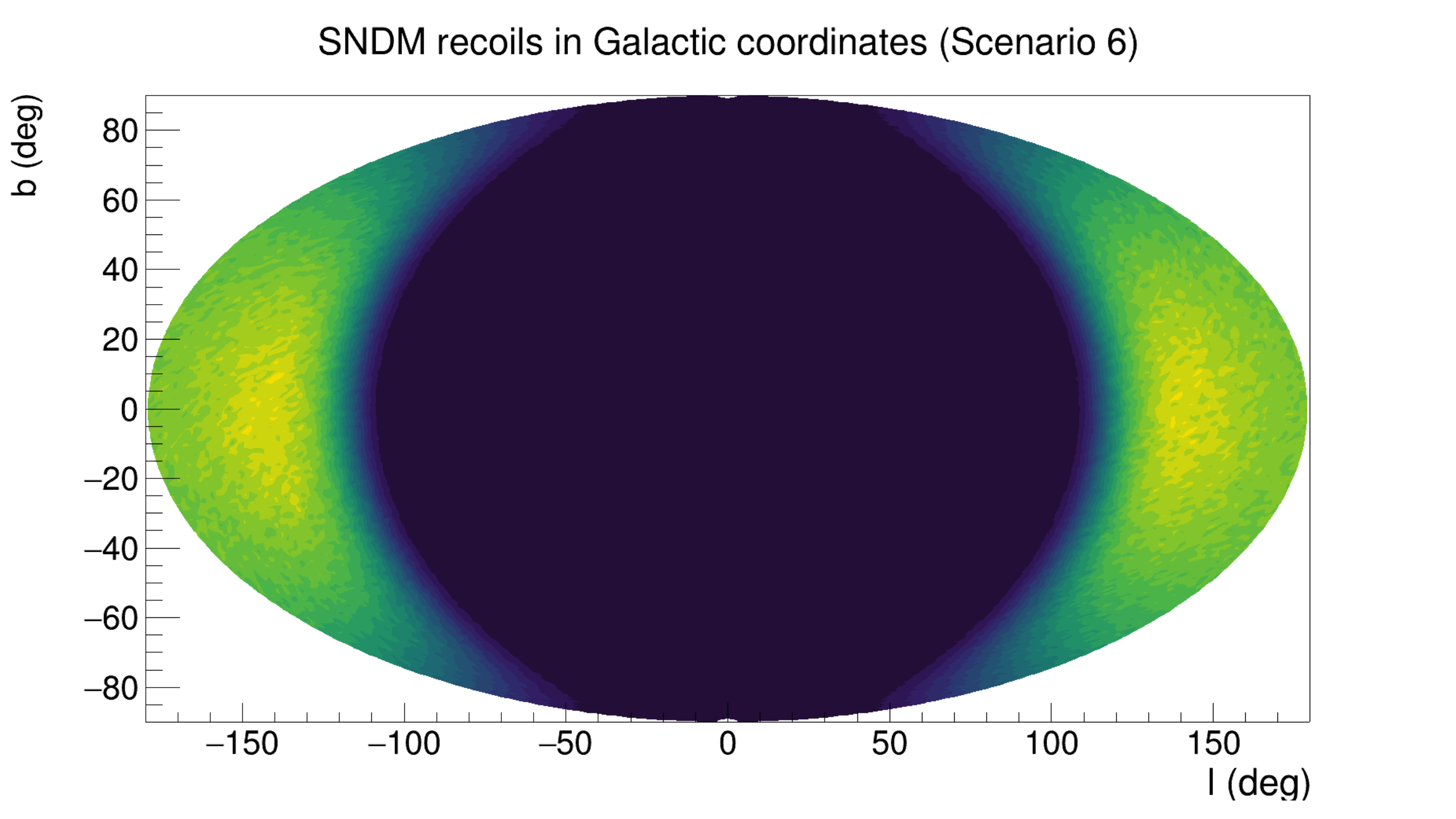}
 \caption{Comparison of the angular distribution of nuclear recoils in Galactic coordinates from the WIMP signal (left) and SNDM signal (right) for the six scenarios considered (1 to 6 from top to bottom), where the color scale indicates the recoil density. It is clear that the two models have dramatically different recoil spectra by virtue of their near perpendicular arrival direction.}
 \label{fig:spectra2D}
 \end{figure*}

In Fig.~\ref{fig:spectra2D}, we show the 2D spectrum of nuclear recoils in our fiducial experimental setup for the various comparison scenarios considered (Table~\ref{tab:sn_wimp_cases}). It is clear there is a dramatic difference in average recoil direction by virtue of the approximately perpendicular arrival directions of the SNDM and WIMP signal. This allows powerful discrimination in the 2D case even with fairly low angular resolution.

\section{Statistical test}\label{app:stat}

In this section we will discuss how we estimate the number of signal events needed in order to distinguish the SNDM signal from the WIMP signal, assuming only one measurable quantity is available (either the energy spectrum, 1D angular spectrum, or 2D angular spectrum). 

To begin, note that for the sake of simplicity, we do not perform a joint analysis using both energetic and angular information simultaneously, but treat each as if it is the only spectrum known. Employing the kinematic correlation between energy and angle in experiments that can infer track direction would only further improve an experiment's capability to discriminate between various models.
Moreover, it is important to reiterate that the study has been performed in the assumption of no background, i.e. that the energy thresholds chosen guarantee full background rejection in the ROI. As discussed in Sec.~\ref{sec:toy}, the angular distribution of background events is expected to be isotropic and therefore to significantly differ from both SNDM or WIMP angular spectra, while still resembling their energy spectrum. 

The statistical technique we use as our test is based on a profile likelihood ratio test, by which we find the number of detected events $\mu_n$ in either the energy, 1D angle, or 2D angle spectra needed to discriminate between two different hypothetical DM signals (SNDM and WIMP).  
In order to do this, we randomly extract $\mu_n$ events according to the energy and angular spectra defined in Sec.~\ref{sec:kinematic} and shown in Fig.~\ref{fig:spectra}. With these, we fill a histogram to represent the outcome of the measurement performed by an experiment, after a proper Gaussian smearing of the extracted quantity following the expected experimental resolutions illustrated in Sec.~\ref{sec:expparameter}. The range of the histogram allows us to take into account the ROI discussed in the Sec.~\ref{sec:expparameter} and the bin sizes are chosen to be twice the $\sigma$ resolution evaluated in the centre of the bin itself (i.e. if $x_0$ is the centre of the bin, the bin range will go from $x_0-\sigma_{res}(x_0)$ to $x_0+\sigma_{res}(x_0)$). 

Given these distributions, the likelihoods under two hypotheses are calculated: WIMP signal (hypothesis $H_0$) and SNDM signal (hypothesis $H_1$). The likelihood is evaluated as a simple multinomial PDF as follows:
\be
\label{eq:likelihood}
 L_{y|x}=\mu_{n}!\prod_{i=1}^{N_{\text{bins}}} \left[ \left( \sum_{j=i}^{N_{\text{adjacent}}} P^{\text{migrate}}_{j\rightarrow i}P_{j|x} \right)^{n_i}\frac{1}{n_i!}\right]
\ee
where
\begin{itemize}
	\item $ x$ denotes the assumed hypothesis: H$_0$ (WIMP) or H$_1$ (SNDM)
	\item $ y$ denotes the Monte Carlo data generated for each model from the analytic spectral formulae: $W$ (WIMP) or $S$ (SNDM)
	\item $ n_i $ is the number of signal events in the $i$th bin
	\item the product runs over all the bins of the histogram of the experiment (the term becomes irrelevant if $ n_i=0 $)
	\item $ P_{j|x} $ is the probability of an event in the $j$th bin under hypothesis $x$
	\item $P^{\text{migrate}}_{j\rightarrow i} $ is the probability of an event that occurred in the $j$th bin migrates to the $i$th bin due to resolution effects, which captures the effect of spectral smearing due to imperfect resolution
	\item the sum runs over all the bins adjacent to the $i$th one ($i$th included) in the histogram. (Note: for the energy histograms both adjacent and next-to-adjacent bins are considered, corresponding to a total of 5 bins are in the sum; for the 1D angular histograms only truly adjacent bins are considered, corresponding to 3 bins in the 1D case and 9 bins in the 2D case).
\end{itemize}

The evaluation of $ P^{\text{migrate}}_{j\rightarrow i} $ is performed under the approximation that when an event migrates to the $j$th bin its value is shifted to the center of that bin. Thus, for example, under our assumption of $1\sigma$ bin width and Gaussian smearing, $ P^{\text{migrate}}_{i\rightarrow i}=0.683 $ and $ P^{\text{migrate}}_{i+1\rightarrow i}=0.159 $ in the case of the 1D angle. (The other cases are computed similarly for the five bins in energy and nine bins in 2D angle.)

From these assumptions, it is possible to construct the likelihood ratio $LR$ defined as
\be
\label{eq:likeliratio}
	LR_{y}=\frac{L_{y|H_1}}{L_{y|H_0}}
\ee
with $y=W,S$. We generate 5$\cdot$10$^5$ Monte Carlo experiments to obtain the probability distributions of the likelihood ratios $f_W(LR)$ and $f_S(LR)$. 
We state that $\mu_n$ events are enough to distinguish between the two models if both the probabilities of committing either a type I or type II error are 5$\%$ or less to ensure symmetry between the two hypotheses.
(Note that $\mu_n$ is never chosen smaller than 3, as this is the minimum number of events needed to distinguish a Poisson fluctuation from zero background.)

\section{Tables of results}
\label{app:tables}

This section simply presents the results shown in Fig.~\ref{fig:relall} in tabular form.

\begin{table*}[h!]

\centering
\begin{center}

\begin{center}

\begin{tabular}{c|c|c|c|c|c|c}
Scenario & Target & WIMP Mass [GeV]  & SNDM Mass [MeV]  & $N_{\text{events}}$ (Energy) & $N_{\text{events}}$ (1D Angle)  & $N_{\text{events}}$ (2D Angle) \\ 
\hline
\hline
1 &$^4$He & 10  & 5& 65 & 3 & 3 \\ 
2 & $^{19}$F & 10 & 7 & 105 & 3 & 3 \\ 
3 & $^{131}$Xe & 10 & 9 & 345 & 3 & 3 \\ 
4 & $^4$He & 100 & 5 & 43 & 4 & 3 \\ 
5 & $^{19}$F & 100 & 14 & 28  & 6 & 3 \\ 
6 & $^{131}$Xe & 100 & 38 & 109 & 21 & 3
\end{tabular}
\caption{Number of events needed to discriminate the two models in the fiducial experimental setup with different spectral information and $2^{\circ}$ angular resolution.}
\label{tab:results2}
\end{center}

\centering
\begin{center}
\begin{tabular}{c|c|c|c|c|c|c}
Scenario &Target & WIMP Mass [GeV]  & SNDM Mass [MeV]  & $N_{\text{events}}$ (Energy) & $N_{\text{events}}$ (1D Angle)  & $N_{\text{events}}$ (2D Angle) \\ 
\hline
\hline
1 & $^4$He & 10  & 5& 65 & 3 & 3 \\ 
2 & $^{19}$F & 10 & 7 & 105 & 4 & 3 \\ 
3 & $^{131}$Xe & 10 & 9 & 345 & 3 & 3 \\ 
4 & $^4$He & 100 & 5 & 43 & 4 & 3 \\ 
5 & $^{19}$F & 100 & 14 & 28  & 8 & 3 \\
6 & $^{131}$Xe & 100 & 38 & 109 & 29 & 4
\end{tabular}
\caption{Number of events needed to discriminate the two models in the fiducial experimental setup with different spectral information and $10^{\circ}$ angular resolution.}
\label{tab:results10}
\end{center}

\centering
\begin{center}
\begin{tabular}{c|c|c|c|c|c|c}
Scenario &Target & WIMP Mass [GeV]  & SNDM Mass [MeV]  & $N_{\text{events}}$ (Energy) & $N_{\text{events}}$ (1D Angle)  & $N_{\text{events}}$ (2D Angle) \\ 
\hline
\hline
1 & $^4$He & 10  & 5& 65 & 7 & 3 \\ 
2 & $^{19}$F & 10 & 7 & 105 & 7 & 3 \\ 
3 & $^{131}$Xe & 10 & 9 & 345 & 5 & 3 \\ 
4 & $^4$He & 100 & 5 & 43 & 9 & 3 \\ 
5 & $^{19}$F & 100 & 14 & 28 & 16 & 4 \\
6 & $^{131}$Xe & 100 & 38 & 109 & 45 & 5
\end{tabular}
\caption{Number of events needed to discriminate the two models in the fiducial experimental setup with different spectral information and $20^{\circ}$ angular resolution.}
\label{tab:results20}
\end{center}

\centering
\begin{center}
\begin{tabular}{c|c|c|c|c|c|c}
Scenario & Target & WIMP Mass [GeV]  & SNDM Mass [MeV]  & $N_{\text{events}}$ (Energy) & $N_{\text{events}}$ (1D Angle)  & $N_{\text{events}}$ (2D Angle) \\ 
\hline
\hline
1 & $^4$He & 10  & 5& 65 & 13 & 4 \\ 
2 & $^{19}$F & 10 & 7 & 105 & 15 & 4 \\ 
3 & $^{131}$Xe & 10 & 9 & 345 & 9 & 3 \\ 
4 & $^4$He & 100 & 5 & 43 & 17 & 4 \\ 
5 & $^{19}$F & 100 & 14 & 28  & 27 & 5 \\
6 & $^{131}$Xe & 100 & 38 & 109 & 76 & 8
\end{tabular}
\caption{Number of events needed to discriminate the two models in the fiducial experimental setup with different spectral information and $30^{\circ}$ angular resolution.}
\label{tab:results30}
\end{center}

\begin{tabular}{c|c|c|c|c|c|c}
Scenario &Target & WIMP Mass [GeV]  & SNDM Mass [MeV]  & $N_{\text{events}}$ (Energy) & $N_{\text{events}}$ (1D Angle)  & $N_{\text{events}}$ (2D Angle) \\ 
\hline
\hline
1 & $^4$He & 10  & 5& 65 & 29 & 7 \\ 
2 & $^{19}$F & 10 & 7 & 105 & 30 & 9 \\ 
3 & $^{131}$Xe & 10 & 9 & 345 & 20 & 9 \\ 
4 & $^4$He & 100 & 5 & 43 & 4 & 3 \\ 
5 & $^{19}$F & 100 & 14 & 28  & 56 & 11 \\
6 & $^{131}$Xe & 100 & 38 & 109 & 157 & 18
\end{tabular}
\caption{Number of events needed to discriminate the two models in the fiducial experimental setup with different spectral information and $45^{\circ}$ angular resolution.}
\label{tab:results45}
\end{center}

\end{table*}

\end{appendix}

\bibliography{ref}

\end{document}